\documentclass[preprint,aps]{revtex4}
\usepackage{graphicx}
\def\Vec#1{\mbox{\boldmath $#1$}}

\begin{document}

\title{
Dilute Multi Alpha Cluster States in Nuclei
}

\author{Taiichi Yamada}
\address{
\it Laboratory of Physics, Kanto Gakuin University, 
Yokohama 236-8501, Japan
}
\address{
\it Institute de Physique Nucl\'eaire, F-91406 Orsay Cedex, France
}
  
\author{Peter Schuck}
\address{
\it Institute de Physique Nucl\'eaire, F-91406 Orsay Cedex, France
}

\date{\today}

\begin{abstract}
Dilute multi $\alpha$ cluster condensed states with spherical and
 axially deformed shapes are studied with the Gross-Pitaevskii equation 
 and Hill-Wheeler equation, where the $\alpha$ cluster is treated 
 as a structureless boson.
Applications to self-conjugate $4N$ nuclei show that the dilute $N\alpha$ 
 states of $^{12}$C to $^{40}$Ca with $J^\pi=0^+$ appear in the energy region 
 from threshold up to about 20 MeV, and the critical number 
 of $\alpha$ bosons that the dilute $N\alpha$ system can 
 sustain as a self-bound nucleus is estimated roughly to be $N_{cr}\sim10$.
We discuss the characteristics of the dilute $N\alpha$ states 
 with emphasis on the $N$ dependence of their energies and rms radii. 
\end{abstract}

\maketitle

\section{Introduction}
%
The molecular-like picture as well as the single-particle picture are  
 fundamental viewpoints to understand the structure 
 of light nuclei.\cite{Wildermuth77,Brink66,Bertsch71,Fujiwara80}
It is well known that the structure of many states in light nuclei is described successfully by 
 the microscopic cluster model, where a group of nucleons is assumed to form a localized 
 substructure (cluster), interacting with other clusters and/or nucleons in the nucleus.
The predominant cluster is the $\alpha$ nucleus, which plays an important role in the cluster
 model, because it is the lightest and also smallest shell-closed nucleus with a binding energy 
 as large as 28 MeV, reflecting the strong four-nucleon correlation.
Molecular-like states in nuclei are expected to appear around the threshold energy of 
 breakup into constituent clusters \cite{Ikeda68}, because the intercluster binding is weak 
 in the cluster states.
For example, the ground state of $^8$Be and the second $0^+$ state of $^{12}$C
 are known \cite{Horiuchi86} to have loosely bound $2\alpha$ and $3\alpha$ structures, respectively, 
 which appear around the $2\alpha$ and $3\alpha$ thresholds.

Special attention has been paid to such four-nucleon correlations corresponding 
 to an $\alpha$-type condensation in symmetric nuclear matter, 
 similar to the Bose-Einstein condensation for finite number of dilute bosonic atoms such as 
 $^{87}$Rb or $^{23}$Na at very low temperature~\cite{Dalfovo99}. 
Several authors have discussed the possibility of $\alpha$-particle condensation 
 in low-density nuclear matter~\cite{Ropke98,Beyer00}.
They found that such $\alpha$ condensation can occur in the low-density region below 
 a fifth of the saturation density, although the ordinary pairing correlation can prevail 
 at higher density.
The result indicates that $\alpha$ condensate states in finite nuclear system may exist 
 in excited states of dilute density composed of weakly interacting gas of $\alpha$ particles.
Thus, it is an interesting subject to study the structure of light nuclei from the viewpoint of
 $\alpha$-particle condensation.
 
Recently, a new $\alpha$-cluster wave function was proposed which is of the $N\alpha$-particle 
 condensate type \cite{Tohsaki01}:
\begin{eqnarray}
&&|\Phi_{N\alpha}\rangle=(C^+_\alpha)^N |{\rm vac}\rangle,\label{C_dager}\\
&&\langle\Vec{r}_1\cdots\Vec{r}_N|\Phi_{N\alpha}\rangle\propto{\mathcal A}\left\{ e^{-\nu\left(\Vec{r}_1^2+\cdots+\Vec{r}_N^2\right)}
\phi(\alpha_1)\cdots\phi(\alpha_N)\right\},\label{wf}
\end{eqnarray}
 where $C^+_\alpha$ is the $\alpha$-particle creation operator, $\phi(\alpha)$ denotes
 the internal wave function of the $\alpha$ cluster, $\Vec{r}_i$ is the center-of-mass
 coordinate of the {\it i}-th $\alpha$ cluster, and ${\mathcal A}$ presents 
 the antisymmetrization operator among the constituent nucleons.
The important characteristic in Eq.~(\ref{wf}) is that the center-of-mass motion of 
 each $\alpha$ cluster is of $S$-wave type with the independent size parameter $\nu$.
Applications of the condensate-type wave function to $^{12}$C and $^{16}$O showed that 
 the second $0^+$ state of $^{12}$C ($E_x$=7.65 MeV) and fifth $0^+$ states of $^{16}$O
 ($E_x$=14.0 MeV), located around the $3\alpha$- and $4\alpha$-particle thresholds, 
 respectively, are specified by the $N\alpha$ condensate state, which is quite similar 
 to the Bose-Einstein condensation of bosonic atoms in magnetic traps where 
 all atoms occupy the lowest $S$-orbit.\cite{Tohsaki01}
The calculated root-mean-square (rms) radius for 
 those condensate states is about 4 fm, which is much much larger than that 
 for the ground state (about 2.7 fm). 
As for $^8$Be, the $\alpha$-particle wave function in Eqs.~(\ref{C_dager}) and (\ref{wf}), 
 taking into account the axially symmetric deformation, was applied to investigate 
 the rotational structure of the ground-band states ($0^+$-$2^+$-$4^+$) \cite{Funaki02}.
It was found that the rotational character of the $^8$Be ground state band
 is reproduced nicely by this wave function with deformation \cite{Funaki02}.

The above-mentioned theoretical results for $^8$Be, $^{12}$C and $^{16}$O 
 lead us to conjecture that such dilute $\alpha$-cluster states near the $N\alpha$
 threshold may also occur in other heavier $4N$ self-conjugate nuclei.
The Coulomb potential barrier should play an important role to confine
 such dilute $N\alpha$-particle states, as inferred from the analyses of the
 $^8$Be, $^{12}$C and $^{16}$O nuclei. 
In fact, the ground state of $^8$Be, which appears at $E_{2\alpha}=92$ keV 
 referring to the $2\alpha$ threshold, exists as a resonant state with very narrow width,
 due to the Coulomb potential barrier, whose height is estimated to be $1\sim2$ MeV. 
This self-trapping of $\alpha$-particles by the Coulomb barriers is in contrast 
 to the case of the dilute neutral atomic condensate states, where the atoms are 
 trapped by the external magnetic field~\cite{Dalfovo99}.

It is an intriguing problem and of interest to study how many $\alpha$ particles 
 can be bound in the dilute nuclear system.
Increasing the number of $\alpha$ clusters, the rms radius of the system 
 should become gradually larger, because the dilute character must be retained.
Then, the total kinetic energy of the $N\alpha$ system becomes significantly smaller 
 in comparison with the potential energy, similar to the case of bosonic atoms 
 in the condensed state.
In addition, the height of the Coulomb potential barrier will become steadily lower. 
This is due to the following reasons.
The $\alpha$-$\alpha$ nuclear potential is short-ranged, while the Coulomb potential 
 is long-ranged.
Increasing the number of the $\alpha$ particles, the repulsive contribution 
 from the Coulomb potentials prevails gradually over the attractive one from 
 the $\alpha$-$\alpha$ nuclear potentials, because the gas-like 
 $N\alpha$ system is expanding in such a way that the rms radius between 
 two $\alpha$ particles is gradually getting larger.
This means that the height of the potential barrier confining the gas-like $N\alpha$ 
 particles becomes lower with increasing $N$, and then, there should exist a critical 
 number ($N=N_{cr}$) beyond which the $N\alpha$ system can not sustain itself as a bound 
 nuclear state.
Thus, it is interesting to estimate the number of $N_{cr}$ as well as to study
 the structure of the dilute $N\alpha$ system up to $N=N_{cr}$.
Since the application of the condensate-type wave function 
 in Eqs.~(\ref{C_dager}) and (\ref{wf}) to the general $N\alpha$ system is not easy 
 and the computational limitation is $N\sim6$ at most, we need to study them 
 with phenomenological models.  

In this paper, the gas-like $N\alpha$ cluster states are studied
 with the following two approaches: the Gross-Pitaevskii-equation approach and 
 Hill-Wheeler-equation approach.
The Gross-Pitaevskii equation, which is of the nonlinear-Schr\"odinger-type, was 
 proposed about fifty years ago to describe the single-boson motion in the dilute
 atomic condensate state~\cite{Pitaevskii61}.
Recent experiments starting from the middle of 1990's have succeeded in realizing 
 such a condensate state consisting of $10^5$-$10^6$ neutral atoms
 trapped by the magnetic field at very low temperature.
Many characteristic aspects of the dilute states are described successfully 
 by the Gross-Pitaevskii equation~\cite{Dalfovo99}. 
Thus, its application to the dilute $N\alpha$ nuclear systems 
 is also very promising, and we expect from such a study useful information 
 on the condensed states.
On the other hand, we here propose also a different approach.
Although the Gross-Pitaevskii equation is simple and interesting for the study of
 the structure of the dilute $N\alpha$ system, the center-of-mass motion
 in the $N\alpha$ system is not completely removed in this framework.
The effect of the center-of-mass motion should be non-negligible
 to the total energy and rms radius etc.~for small numbers of $\alpha$ bosons.
We, thus, formulate in this paper the framework describing the dilute $N\alpha$ 
 systems free from the center-of-mass motion, with the approach of the 
 Hill-Wheeler equation.
The spherical $N\alpha$ systems as well as the deformed ones, with
 the axial symmetry, are discussed with this equation.
The two different approaches, the Gross-Pitaevskii-equation approach and
 the Hill-Wheeler-equation approach, are complementary to one another,
 and we will obtain useful understanding of the dilute multi $\alpha$-particle states. 
Starting from $^8$Be, the structure of the gas-like $N\alpha$ boson systems
 with $J^\pi=0^+$ is investigated and the critical number $N_{cr}$ is estimated 
 with those two different approaches.

The paper is organized as follows.
We formulate in Sec.~II the Gross-Pitaevskii-equation approach and the
 Hill-Wheeler-equation approach for the dilute $N\alpha$ system.
In Sec.~III the calculated results are presented, and the characteristics 
 of the dilute $N\alpha$ states are discussed with emphasize on the $N$ dependence 
 of their energies and rms radii.
A summary finally is given in Sec.~IV. 

\section{Formulation}
%
In this section, we formulate the two approaches to study 
 the structure of the dilute $N\alpha$ nuclear systems: the Gross-Pitaevskii-equation
 approach and Hill-Wheeler-equation approach.

\subsection{Gross-Pitaevskii equation for dilute $N\alpha$ nuclear systems}
%
In the mean field approach, the total wave function of the condensate $N\alpha$-boson system 
 is represented as
\begin{eqnarray}
\Phi(N\alpha)=\prod_{i=1}^{N}\varphi(\Vec{r}_i),
\end{eqnarray}
where $\varphi$ and $\Vec{r}_i$ are the normalized single-$\alpha$ wave function and
 coordinate of the {\it i}-th $\alpha$ boson.
Then, the equation of motion for the $\alpha$ boson, called as the Gross-Pitavskii equation,
 is of non-linear Schr\"odinger-type,  
\begin{eqnarray}
&&-\frac{\hbar^2}{2m}\left(1-\frac{1}{N}\right)\Vec{\nabla}^2\varphi(\Vec{r})
   +U(\Vec{r})\varphi(\Vec{r})=\varepsilon\varphi(\Vec{r}),\label{GPE}\\
&&U(\Vec{r})=(N-1)\int d\Vec{r}'\left|\varphi(\Vec{r}')\right|^2{\upsilon_2(\Vec{r}',\Vec{r})}\nonumber\\
&&\hspace*{1cm}+\frac{1}{2}(N-1)(N-2)\int d\Vec{r}''d\Vec{r}'
  {\left|\varphi(\Vec{r}'')\right|^2}{\left|\varphi(\Vec{r}')\right|^2}{\upsilon_3(\Vec{r}'',\Vec{r}',\Vec{r})},
  \label{GPE_alpha_pot}
\end{eqnarray}
where $m$ stands for the mass of the $\alpha$ particle, $U$ is the mean-field potential of
 $\alpha$-particles, and $\upsilon_2$ ($\upsilon_3$) denotes the $2\alpha$ ($3\alpha$) interaction.
The center-of-mass kinetic energy correction, $1-1/N$, is taken into account together 
 with the finite number corrections, $N-1$, etc.
In the present study, only the $S$-wave state is solved self-consistently with the iterative method.
The total energy of the $N\alpha$ system $E(N\alpha)$ is presented as
\begin{eqnarray}
&&E(N\alpha)=N\left[{\langle t\rangle}+\frac{1}{2}(N-1){\langle \upsilon_2\rangle}+\frac{1}{6}(N-1)(N-2){\langle \upsilon_3\rangle}\right],\\
&&{\langle t\rangle}=\left(1-\frac{1}{N}\right)\times{\langle \varphi(\Vec{r}) |-\frac{\hbar^2}{2m}\Vec{\nabla}^2|
    \varphi(\Vec{r})\rangle},\\
&&{\langle \upsilon_2\rangle}={\langle \varphi(\Vec{r})\varphi(\Vec{r}')|\upsilon_2(\Vec{r},\Vec{r}')|
    \varphi(\Vec{r})\varphi(\Vec{r}')\rangle},\\
&&{\langle \upsilon_3\rangle}={\langle \varphi(\Vec{r})\varphi(\Vec{r}')\varphi(\Vec{r}'')|
    \upsilon_3(\Vec{r},\Vec{r}',\Vec{r}'')|
    \varphi(\Vec{r})\varphi(\Vec{r}')\varphi(\Vec{r}'')\rangle},
\end{eqnarray}
 and the eigen energy $\varepsilon$ in Eq.~(\ref{GPE}) is given as 
\begin{eqnarray}
 \varepsilon={\langle t\rangle}+(N-1){\langle \upsilon_2\rangle}+\frac{1}{2}(N-1)(N-2){\langle \upsilon_3\rangle}.
\end{eqnarray}
The nuclear rms radius in the $N\alpha$ state is evaluated as,  
\begin{eqnarray}
&&\sqrt{\langle r^2_N\rangle}=\sqrt{\langle r^2_\alpha\rangle_{GP}+1.71^2},\label{rms_N}  \\
&&{\langle r^2_\alpha\rangle_{GP}}=\left(1-\frac{1}{N}\right)\langle\varphi|r^2|\varphi\rangle,\label{rms_N_cor}
\end{eqnarray}
 where we take into account the finite size effect of the $\alpha$ particle and the correction
 of the center-of-mass motion.

\subsection{Hill-Wheeler equation for dilute $N\alpha$ states}
%
The framework of the Gross-Pitaevskii approach is simple and useful for
 the study of the structure of the dilute $N\alpha$ systems.
The center-of-mass motion, however, is not removed exactly in this approach.
The effect on the total energy and rms radius etc.~is not negligible for the 
 small-number $\alpha$-boson systems.
We formulate here the Hill-Wheeler-equation approach for the axially symmetric 
 $N\alpha$ systems as well as for the spherical ones,  in which the center-of-mass 
 motion is completely eliminated. 
Only the Hill-Wheeler equation for the deformed $N\alpha$ systems with the
 axial symmetry is provided in this section, because the limit of the spherical
 systems is included in the deformed case.

The single $\alpha$-particle wave function should be mainly in the lowest $S$ 
 state in the gas-like $N\alpha$ boson system.
Thus, the model wave function of the axially symmetric $N\alpha$ system should be 
 described in terms of a superposition of a Gaussian basis with axially
 symmetric deformation (taking the $z$-axis as the symmetric axis) as follows,
\begin{eqnarray}
&&\Phi^{(int)}(N\alpha)=\sum_{\nu_1,\nu_3}f(\nu_1,\nu_3)\Phi^{(int)}(\nu_1,\nu_3),\label{total_wf}\\
&&\Phi^{(int)}(\nu_1,\nu_3)=\int d\Vec{R}_{cm}{{\Phi^{(cm)}}^*(\Vec{R}_{cm}; \nu_1,\nu_3)}\Phi(\nu_1,\nu_3),\\
&&\Phi(\nu_1,\nu_3)=\prod_{i=1}^N\phi(\Vec{r}_i; \nu_1, \nu_3)
                       =\Phi^{(int)}(\nu_1,\nu_3)\Phi^{(cm)}(\Vec{R}_{cm}; \nu_1,\nu_3),\label{cm-rel}\\
&&\phi(\Vec{r}; \nu_1, \nu_3)
 ={\left(\frac{2\pi}{\nu_1}\right)^{\frac{1}{2}}}{\left(\frac{2\pi}{\nu_3}\right)^{\frac{1}{4}}}
 \exp\left(-\nu_1x^2-\nu_1y^2-\nu_3z^2\right),\\
&&\Phi^{(cm)}(\Vec{R}_{cm}; \nu_1,\nu_3)=\phi(\Vec{R}_{cm}; N\nu_1,N\nu_3),
\end{eqnarray}
 where $\Vec{r}_i$ ($\Vec{R}_{cm}$) denotes the coordinate of the {\it i}-th $\alpha$ 
 boson (the center-of-mass coordinate of the $N\alpha$ system),
 and $\nu_1$ ($\nu_3$) presents the Gaussian size parameter of the $x$ and $y$ 
 directions ($z$).
The wave function in Eq.~(\ref{total_wf}) is totally symmetric under any exchange of
 two $\alpha$ bosons.
The angular-momentum-projected total wave function for dilute $N\alpha$ systems free from
 the center-of-mass motion is given as
\begin{eqnarray}
&&\Phi^{(int)}_{JM}(N\alpha)=\sum_{\nu_1,\nu_3}f_J(\nu_1,\nu_3)\Phi^{(int)}_{JM}(\nu_1,\nu_3), 
             \label{projected_wf}\\
&&\Phi^{(int)}_{JM}(\nu_1,\nu_3)=\int {d \cos\theta} {d^J_{M0}(\theta)} {{\mathcal R}_y(\theta)}
             \Phi^{(int)}(\nu_1,\nu_3),\label{J_projected_basis}
\end{eqnarray}
where ${\mathcal R}_y$ denotes the rotation operator around the {\it y} axis.
Setting $\nu_3=\nu_1$ in Eq.~(\ref{projected_wf}), we obtain the wave function
 for the spherical $N\alpha$ system, where only $J^\pi=0^+$ state is allowed.
The matrix element of a translational invariant scalar operator $\hat{O}$ with respect to the 
 angular momentum projected $N\alpha$ wave function in Eq.~(\ref{J_projected_basis}) is
 evaluated as
\begin{eqnarray}
&&\langle\Phi^{(int)}_{JM=0}(\nu_1,\nu_3)|\hat{O}|
  \Phi^{(int)}_{JM=0}({\nu_1}',{\nu_3}')\rangle\nonumber\\
&& \hspace{0.5cm} =\int {d\cos\theta}{d^J_{00}(\theta)}
  \langle\Phi^{(int)}(\nu_1,\nu_3)|{\mathcal R}_y(\theta)\hat{O}|\Phi^{(int)}({\nu_1}',{\nu_3}')\rangle,\\
&& \hspace{0.5cm} =\int {d\cos\theta}{d^J_{00}(\theta)}
  \frac{\langle\Phi(\nu_1,\nu_3)|{\mathcal R}_y(\theta)\hat{O}|\Phi({\nu_1}',{\nu_3}')\rangle}
 {\langle\Phi^{(cm)}(\Vec{R}_{cm};\nu_1,\nu_3)|{\mathcal R}_y(\theta)|
  \Phi^{(cm)}(\Vec{R}_{cm}; {\nu_1}',{\nu_3}')\rangle},
\end{eqnarray}

The total Hamiltonian of the $N\alpha$-boson system is given as 
\begin{eqnarray}
  {\mathcal H}=\sum_{i=1}^{N} t_i -T_{cm}
  +\sum_{i<j} \upsilon_2(\Vec{r}_i,\Vec{r}_j)
  +\sum_{i<j<k} \upsilon_{3}(\Vec{r}_i,\Vec{r}_j,\Vec{r}_k), \label{HW_hamiltonian}
\end{eqnarray}
 where $\upsilon_2$ and $\upsilon_3$ denotes, respectively, the $2\alpha$ and $3\alpha$ interactions.
The kinetic energy of the center-of-mass motion ($T_{cm}$) is subtracted 
 from the Hamiltonian.

The equation of motion for the dilute $N\alpha$-boson states are given 
 in terms of the Hill-Wheeler equation \cite{Hill53}:
\begin{eqnarray}
\sum_{{\nu_1}',{\nu_3}'}\left\{
 \langle\Phi^{(int)}_J(\nu_1,\nu_3)|{\mathcal H}-E|\Phi^{(int)}_J({\nu_1}',{\nu_3}')\rangle\right\} 
 f_J({\nu_1}',{\nu_3}')=0,\label{Hill_Wheeler}
\end{eqnarray}
where ${\mathcal H}$ is the total Hamiltonian of the $N\alpha$-boson system 
 in Eq.~(\ref{HW_hamiltonian}).
The coefficients $f_J$ and eigen energies $E$ are obtained by solving the Hill-Wheeler equation. 
In the numerical calculation, the Gaussian size parameters $\nu_1$ and $\nu_3$ are
 discretized and chosen to be of geometric progression, 
\begin{eqnarray}
 &&\nu_1^{(k)}=\left(1/b_1^{(k)}\right)^2,\hspace{0.5cm}b_1^{(k)}=b_1^{(1)}r_1^{k-1},
 \hspace{0.5cm}{k=1\sim k_{max}},\label{gs_para_1}\\
 &&\nu_3^{(K)}=\left(1/b_3^{(K)}\right)^2,\hspace{0.5cm}b_3^{(K)}=b_3^{(1)}r_3^{K-1},
 \hspace{0.5cm}{K=1\sim K_{max}},\label{gs_para_3}
\end{eqnarray}
The above choice of the Gaussian range parameters is found to be suitable for describing 
 the dilute $N\alpha$ states. 

 The nuclear rms radius measured from the center-of-mass coordinate 
 in the $N\alpha$ state, is expressed as, 
\begin{eqnarray}
&&\sqrt{\langle r^2_N\rangle}=\sqrt{\langle r^2_\alpha\rangle_{HW}+1.71^2},\label{HW_rms_N}  \\
&&{\langle r^2_\alpha\rangle}_{HW}
=\langle\Phi^{(int)}_J(N\alpha)|\frac{1}{N}\sum_{i=1}^{N}(\Vec{r}_i-\Vec{R}_{cm})^2|
          \Phi^{(int)}_J(N\alpha)\rangle,\label{rms_alpha}
\end{eqnarray}
 where we take into account the finite size effect of the $\alpha$ particle.
The rms distance between two $\alpha$ particles is given as 
\begin{eqnarray}
\sqrt{\langle r^2_{\alpha\alpha}\rangle}={\left\langle\frac{1}{N(N-1)}\sum_{i,j}(\Vec{r}_i-\Vec{r}_j)^2\right\rangle}^{1/2}
=\left(\frac{2N}{N-1}\right)^{1/2}\times\sqrt{\langle r^2_\alpha\rangle}_{HW}.
\end{eqnarray}
Thus, it is proportional to the rms radius of an $\alpha$-particle from the center-of-mass coordinate.

\subsection{Effective $\alpha$-$\alpha$ potentials.}
%
In the present paper, we use two kinds of effective potentials: the density-dependent potential
 and phenomenological $2\alpha$ plus $3\alpha$ potential.
They are applied to the Gross-Pitaevskii equation and the Hill-Wheeler equation for the study
 of the structure of dilute $N\alpha$ states.

\subsubsection{Density-dependent potential}
The density dependent potential consists of the Gaussian-type $\alpha$-$\alpha$ potential
 including a density-dependent term, which is of similar form as the Gogny potential 
 (known as an effective $NN$ potential) used in nuclear mean-field calculations,
\begin{eqnarray}
&&\upsilon_2(\Vec{r},\Vec{r}')=\upsilon_{0}\exp\left[-0.7^2(\Vec{r}-\Vec{r}')^2\right]
     -130\exp\left[-0.475^2(\Vec{r}-\Vec{r}')^2\right]\nonumber\\
&&\hspace*{1.5cm}+(4\pi)^2g\delta(\Vec{r}-\Vec{r}')\rho\left(\frac{\Vec{r}+\Vec{r}'}{2}\right)
     +\upsilon_{Coul}(\Vec{r},\Vec{r}'),
\label{DD_pot} 
\end{eqnarray}
 where the units of $\upsilon_2$ and $r$ are MeV and fm, respectively, and $\rho$ denotes
 the density of the $N\alpha$ system.
The folded Coulomb potential $\upsilon_{Coul}$ is presented as 
\begin{eqnarray}
\upsilon_{Coul}(\Vec{r},\Vec{r}')=\frac{4e^2}{|\Vec{r}-\Vec{r}'|}{\rm erf}(a|\Vec{r}-\Vec{r}'|).
\label{Coulomb_pot}
\end{eqnarray}
The Gaussian-potential part in Eq.~(\ref{DD_pot}) is based on the Ali-Bodmer potential \cite{Ali66}, 
 which is known to reproduce well the elastic $\alpha$-$\alpha$ scattering phase shift 
 up to about 60 MeV for $\upsilon_0=500$ MeV.
On the contrary, here, the two parameters, $\upsilon_0$ and $g$, are chosen so as to reproduce 
 well the experimental energy 
 ($E^{exp}_3$=0.38 MeV) and the calculated rms radius (4.29 fm by Tohsaki et al. \cite{Tohsaki01}) 
 for the $0_2^+$ state of $^{12}$C by solving the Gross-Pitaevskii equation in Eq.~(\ref{GPE}).
The results are $\upsilon_0$=271 MeV and $g$=1650 MeV$\cdot$fm$^6$, where the calculated
 energy and rms radius for the $3\alpha$ system are 0.38 MeV and 4.14 fm, respectively,
 which are discussed in Sec.~III.
Although the choice of the phenomenological potential in Eq.~(\ref{DD_pot}) is rather rough, 
 it is interesting to study systematically the structure of the $N\alpha$ condensate states
 as a function of $N$.

\subsubsection{Phenomenological $2\alpha$ and $3\alpha$ potential}
There are many phenomenological $2\alpha$ potentials proposed so far.
Since the $\alpha$ particle is treated as a point-like boson in the present study,
 we will use a $2\alpha$ potential taking into account the Pauli
 blocking effect.
The typical potential is the Ali-Bodmer one \cite{Ali66}, which is used frequently 
 in the structure calculation;
 $\upsilon_2(\Vec{r},\Vec{r}')=500 \exp\left[-0.7^2(\Vec{r}-\Vec{r}')^2\right]
                      -130 \exp\left[-0.475^2(\Vec{r}-\Vec{r}')^2\right]
 + \upsilon_{Coul}(\Vec{r},\Vec{r}'),$
where $\upsilon_{Coul}$ denotes the folded Coulomb potential given in Eq.~(\ref{Coulomb_pot}).
The strong repulsion in the inner region prevents the $2\alpha$ particles
 from approaching one another.
It is, however, found that the potential is not suitable to describe the property 
 of the compact shell-model-like structure of the $^{12}$C ground state 
 with the $3\alpha$ boson model~\cite{Tamagaki77}.
However, we may use it for the dilute $N\alpha$ states.
The Ali-Bodmer potential, however, has the following three unfavorable 
 properties for the present calculation.

First is that the potential does not give the experimental resonant energy 
 of the $^8$Be ground state 
 ($E^{cal}_{2\alpha}$=68 keV vs. $E^{exp}_{2\alpha}$=92 keV ), 
 although the $\alpha$-$\alpha$ scattering phase shift is reproduced 
 nicely up to about 60 MeV.
The second is that applying the potential to the $3\alpha$ boson system 
 the lowest energy state obtained corresponds to a relatively compact 
 $3\alpha$-structure state, although the condensate state appears around 
 the $3\alpha$ threshold.
According to the stochastic variational calculation \cite{Suzuki02}, for example,
 the calculated energy and rms radius are, respectively, 
 $E^{cal}_{3\alpha}=-0.62$ MeV and 
 $\sqrt{\langle r^2_N\rangle}$=3.15 fm \cite{footnote}.
The results indicate that the Ali-Bodmer potential is not adequate for describing
 the dilute $3\alpha$-boson state of $^{12}$C.
The third unfavorable point comes from the fact that the strong repulsive 
 character in the short-range region of the Ali-Bodmer potential
 ($\sim400$ MeV) leads us to treat exactly the short-range correlation 
 between the $2\alpha$ bosons.
The treatment is difficult and needs time-consuming numerical calculations 
 for solving general $N\alpha$-boson systems, even if employing modern 
 numerical methods for many body systems.   

\begin{figure}
\begin{center}
\includegraphics*[scale=0.5]{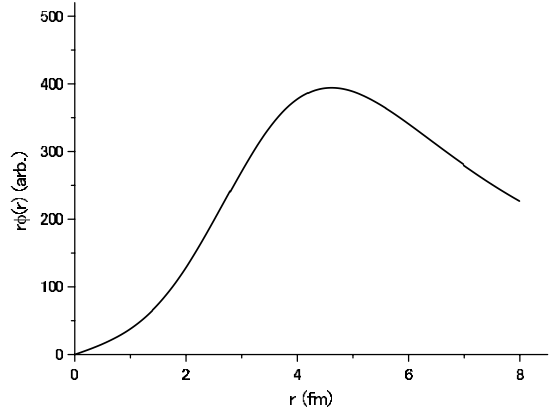}
\caption{
Radial part of the relative wave function between the $2\alpha$ 
clusters in the $J^\pi=0^+$ resonant state at $E_{2\alpha}$=92 keV
with use of the soft-core $\alpha$-$\alpha$ potential 
in Eq.~({\ref{2_body}}).
The scale of $r\phi(r)$ is arbitrary.
}
\label{fig:1}
\end{center}
\end{figure}

We construct here an effective $2\alpha$ potential 
 with a weak repulsive part (soft core), which overcomes the 
 above-mentioned three unfavorable properties.
The use of such a soft-core-type $2\alpha$ potential is suitable
 in the present study, because we discuss the gas-like $N\alpha$ states,
 where the Pauli blocking effect between two $\alpha$ particles is 
 considerably weakened.
The important point for the determination of the potential parameters, other than
 the condition of reproducing the experimental resonant energy for 
 the $^8$Be ground state, is that the $\alpha$-$\alpha$ wave function 
 for the resonant state, should have a loosely-bound structure of the two 
 $\alpha$ particles.
According to many structure calculations of {$^8$Be}, the amplitude of 
 the radial part of the $\alpha$-$\alpha$ relative wave function must be 
 small in the inner region and have a maximum value around $r=4$ fm 
 (where $\Vec{r}$ is the relative coordinate between the two $\alpha$ particles).
This condition ensures that the ground state of $^8$Be has a dilute $2\alpha$ structure.
With a careful search of the potential parameters, we determined the effective 
 $2\alpha$ potential as follows,
\begin{eqnarray}
\upsilon_2(\Vec{r},\Vec{r}')=50\exp\left[-0.4^2(\Vec{r}-\Vec{r}')^2\right]
       -34.101\exp\left[-0.3^2(\Vec{r}-\Vec{r}')^2\right] + \upsilon_{Coul}(\Vec{r},\Vec{r}'),
\label{2_body}
\end{eqnarray}
 where the units of $\upsilon_2$ and $r$ are MeV and fm, respectively,
 and $\upsilon_{Coul}$ is the folded Coulomb potential.
The calculated resonant energy of $^8$Be is $E_{2\alpha}$=92 keV, 
 in agreement with the experimental data ($E^{exp}_{2\alpha}$=92 keV).
Figure~\ref{fig:1} shows the radial part of the relative wave function 
 between the $2\alpha$ clusters for the resonant state.
We see that the amplitude is relatively small for $r=0\sim2$ fm, and 
 has a maximum value around $r$=4 fm.
In spite of the fact that the Ali-Bodmer potential gives a wave function which 
 is almost zero for $r=0\sim1$ fm, the small but finite amplitude around $r=0\sim2$ fm
 should hardly give any effect 
 for dilute multi $\alpha$ cluster states.
Applying this effective potential to the $3\alpha$- and $4\alpha$-boson systems 
 indicates that we get the desired dilute $3\alpha$- and $4\alpha$-structure states 
 for $^{12}$C and $^{16}$O, respectively, within our framework, as shown below. 

In the present study, we introduce the phenomenological $3\alpha$ 
 potential ($\upsilon_{3}$) with repulsive character, 
 as given in Ref.~\cite{Fukatsu92},
\begin{eqnarray}
\upsilon_{3}(\Vec{r},\Vec{r}',\Vec{r}'')
=151.5~{\exp\left\{-{0.15\left[(\Vec{r}-\Vec{r}')^2
   +(\Vec{r}'-\Vec{r}'')^2+(\Vec{r}''-\Vec{r})^2\right]}\right\}},\label{3_body}
\end{eqnarray}
 where the units of $\upsilon_{3}$ and $r$ are MeV and fm, respectively. 
This potential has been used in the $3\alpha$ and $4\alpha$ 
 orthogonally condition model (OCM \cite{Saito68}) for calculations of $^{12}$C and $^{16}$O 
 so as to reproduce the ground-state energies with respect to the $3\alpha$ 
 and $4\alpha$ thresholds, respectively \cite{Fukatsu92}.
In the model, the Pauli principle is taken into account in the relative 
 wave function between two $\alpha$ particles, and a deep attractive potential
 is used for the $\alpha$-$\alpha$ potential.
Thus, the OCM is able to describe not only the shell-model-like compact states 
 but also the dilute gas-like states.
According to the results, the repulsive $3\alpha$ potential gives a large effect 
 to the ground-state energies of $^{12}$C and $^{16}$O with the compact $N\alpha$ 
 structure, while its effect is very small for dilute $3\alpha$ and $4\alpha$ states. 
A non-negligible effect, however, can be expected in large-number $N\alpha$ 
 dilute systems, if we take into account the fact that the contribution of 
 the binding energy from the $3\alpha$ potential, proportional to $N(N-1)(N-2)/6$, 
 raises strongly with increasing $N$.

The reason of why we introduce the repulsive $3\alpha$ potential is given as follows.
Let us define here the total kinetic energy and two-body potential energy 
 in the dilute $N\alpha$ system as $\langle T\rangle$ and $\langle V_2\rangle$, respectively.
In case of the $^8$Be ground state, the experimental energy is  
 $E={\langle T\rangle}+{\langle V_2\rangle}\sim 0.1$ MeV with respect to the $2\alpha$ threshold,
 where ${\langle T\rangle}$ (${\langle V_2\rangle}$) is positive (negative). 
For an arbitrary dilute $N\alpha$-boson system, the total kinetic energy and two-body
 potential energy are given as ${\langle T\rangle}\sim N-1$ 
 and ${\langle V_2\rangle}\sim N(N-1)/2$, respectively, where the center-of-mass kinetic
 energy is subtracted.
Increasing the number of the $\alpha$ particles, therefore, the potential energy prevails
 over the kinetic energy, and then, the system falls gradually into a collapsed state.
This indicates that something like a density-dependent force with the repulsive
 character is needed to avoid the collapse in the large-number $N\alpha$-boson system.
The present repulsive $3\alpha$ potential in Eq.~(\ref{3_body}) plays a role similar
 to the density-dependent force.  
On the other hand, the density-dependent potential given in Eq.~(\ref{DD_pot}) 
 is also used when solving the Gross-Pitaevskii equation and the Hill-Wheeler equation.
The reason of why the density-dependent potential is introduced there is the same as that
 discussed here.

\section{Results and discussion}
%
\subsection{Application of the Gross-Pitaevskii equation to $N\alpha$ systems}

The Gross-Pitaevskii equation is solved with the two different types 
 of effective $\alpha$-$\alpha$ potentials: 1)~the density-dependent potential 
 [see Eq.~(\ref{DD_pot})] and 2)~the phenomenological $2\alpha$ potential with
 the $3\alpha$ potential [see Eqs.~(\ref{2_body}) and (\ref{3_body})].
First of all, we will discuss the results with the density-dependent potential
 and then those with the phenomenological potentials.

\begin{figure}
\begin{center}
\includegraphics*[scale=0.4]{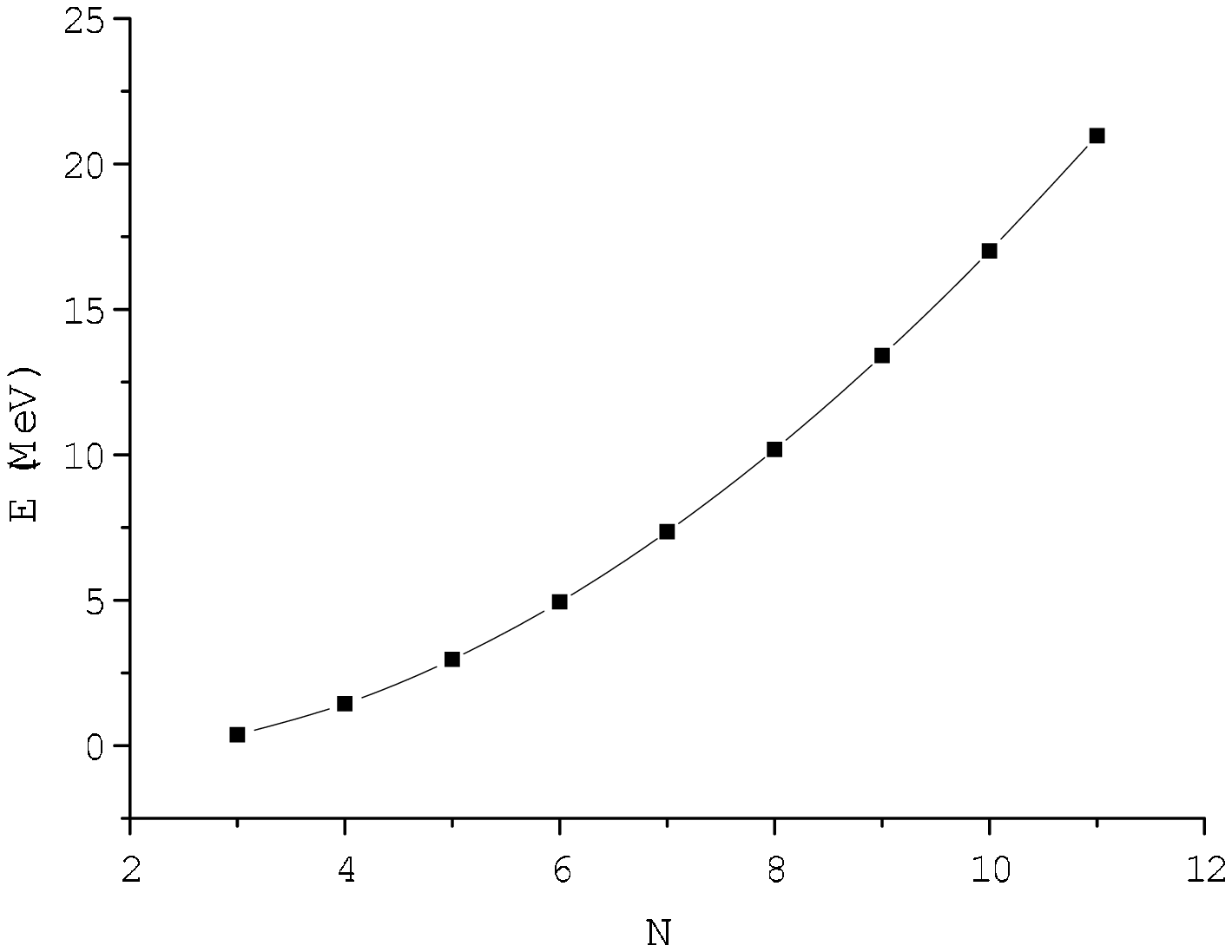}
\includegraphics*[scale=0.4]{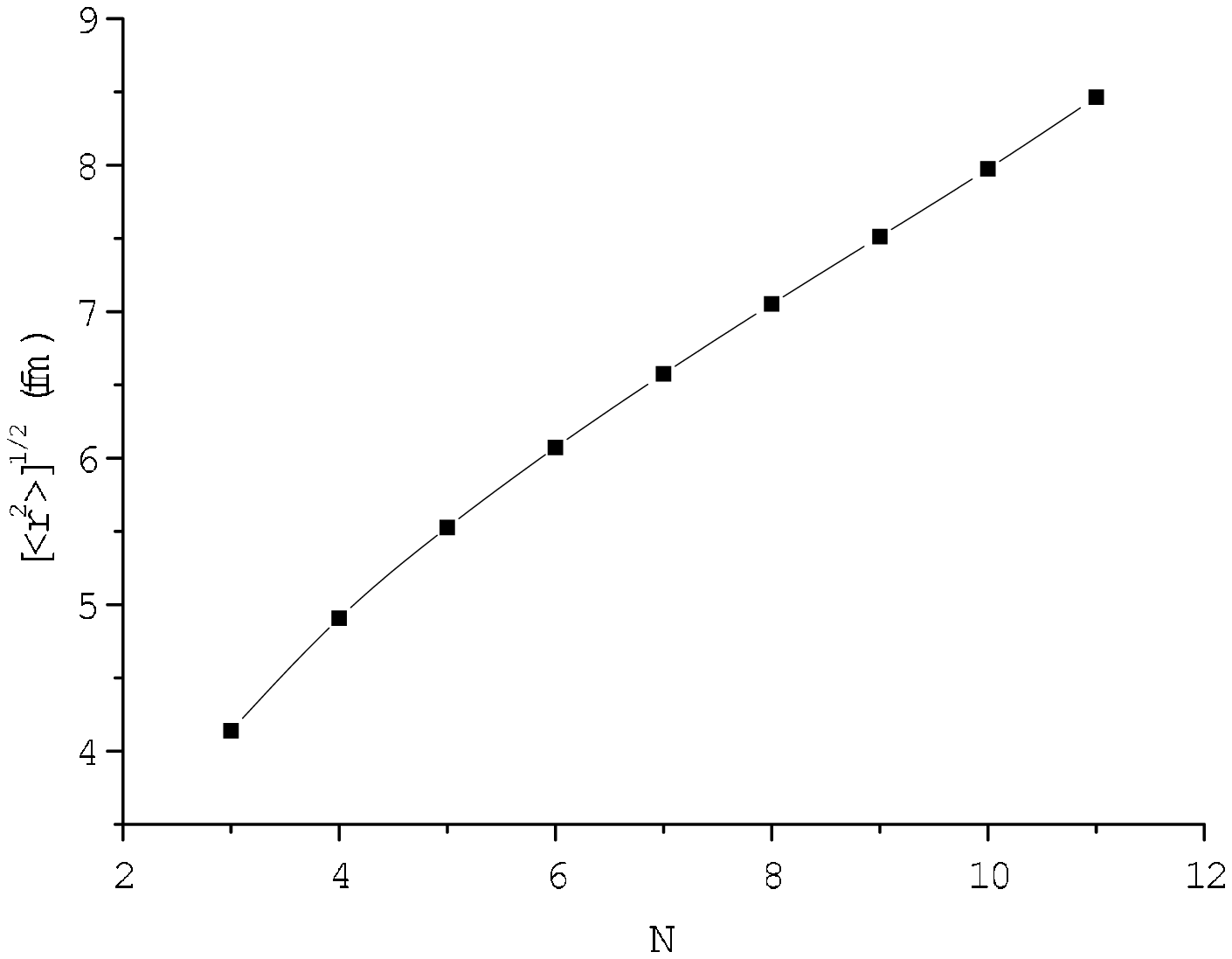}
\caption{
(a)~Total energies for the dilute $N\alpha$ states measured from each $N\alpha$
threshold, and (b)~their nuclear rms radii, which are obtained by solving 
the Gross-Pitaevskii equation with the density-dependent potential. 
}
\label{fig:2}
\end{center}
\end{figure}

The calculated total energies of the $N\alpha$ systems measured from
 the $N\alpha$ threshold are demonstrated in Fig.~\ref{fig:2} 
 as well as the calculated nuclear rms radii defined in Eq.~(\ref{rms_N}), 
 where the density-dependent $\alpha$-$\alpha$ potential is used.
The total energy and the rms radius are getting larger with increasing $N$.
This means that the system is expanding steadily with increase of $N$.
In comparison with the rms radius of the ground state of each 
 nucleus with the empirical formula ($\sqrt{\langle r^2_N\rangle}=1.2A^{1/3}$ fm), 
 the results from the Gross-Pitaevskii equation are much larger
 than those for the ground states.
Thus, the states obtained here can be identified with the dilute $N\alpha$ states.
It is noted that they are obtained naturally from the Gross-Pitaevskii equation with
 the density-dependent $\alpha$-$\alpha$ potential, whose parameters were
 determined so as to reproduce well the experimental energy ($E^{exp}_{3\alpha}$=0.38 MeV)
 and the calculated rms radius by Tohsaki et al.~(4.29 fm)~\cite{Tohsaki01} 
 for the $0^+_2$ state of $^{12}$C.  

\begin{figure}
\begin{center}
\begin{minipage}{0.90\linewidth}
\includegraphics*[width=\linewidth]{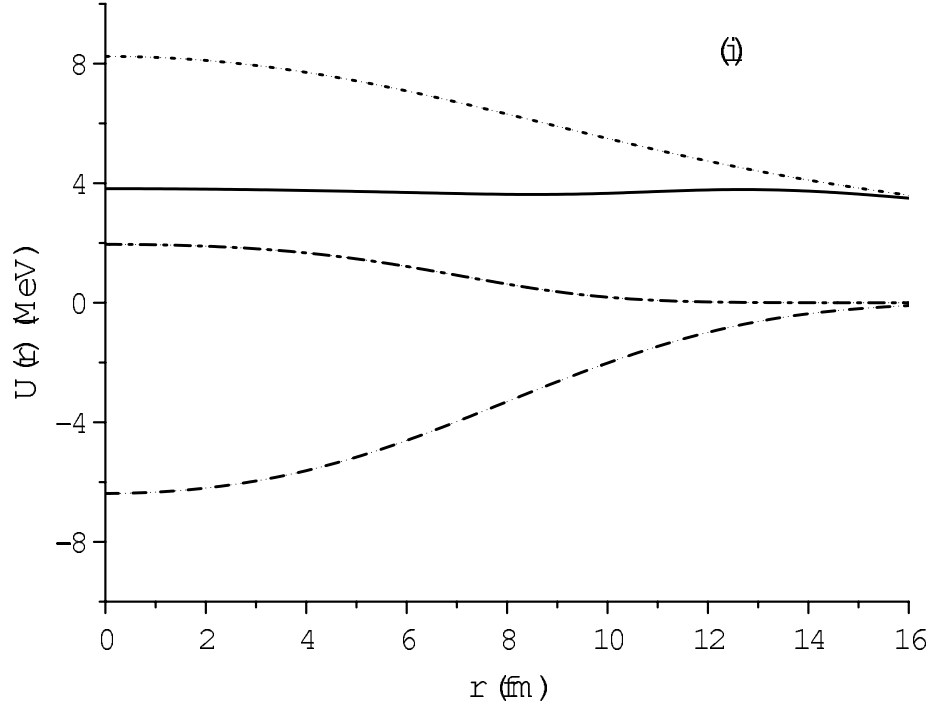}
\end{minipage}
\caption{
Single $\alpha$ particle potentials $U_\alpha(R)$ (solid line) 
which are obtained by solving 
the Gross-Pitaevskii equation with the density-dependent potential;
(a)~$3\alpha$, (b)~$4\alpha$, (c)~$5\alpha$, (d)~$6\alpha$,
(e)~$7\alpha$, (f)~$8\alpha$, (g)~$9\alpha$, (h)~$10\alpha$, 
(i)~$11\alpha$ systems.
The dashed, dot-dashed and dotted lines demonstrate, respectively, the contribution 
from the two-range Gaussian term, density-dependent term and Coulomb potential. 
}
\label{fig:3}
\end{center}
\end{figure}

In order to study the structure of the dilute $N\alpha$ states,
 it is instructive to see the single $\alpha$ potential defined in Eq.~(\ref{GPE_alpha_pot}).
Figures~\ref{fig:3}(a)$\sim$(j) show the ones for the $3\alpha\sim11\alpha$ systems. 
Let us first discuss the $3\alpha$ and $4\alpha$ cases.
The remarkable characteristics of the potentials can be presented as follows:
 1)~the almost flat behavior of the potential in the inside region, and 
 2)~the Coulomb potential barrier in the outer region.
The appearance of the flat potential region is very impressive, if one takes into 
 account the fact that the two-range Gaussian term (attractive), density-dependent 
 term (repulsive) and the Coulomb-potential term (repulsive) contribute significantly
 to the single $\alpha$ potential in the inside region [see Figs.~\ref{fig:3}(a) and (b)].
Table~\ref{tab:1} shows the calculated single $\alpha$ particle energy and 
 contributions from the kinetic energy, two-range Gaussian 
 term in Eq.~(\ref{DD_pot}), density-dependent term in Eq.~(\ref{DD_pot}) 
 and the Coulomb potential. 
It is found that the kinetic energy in the $3\alpha$ and $4\alpha$ systems 
 is not negligible but small in comparison with the two-range-Gaussian 
 term and/or Coulomb potential energy.
This indicates that the Thomas-Fermi approximation, neglecting the kinetic energy term,
 is roughly realized in the system.
Consequently, there appears the flat potential region
 in the single $\alpha$ potential, whose behavior is similar to the dilute 
 atomic condensate state trapped by the magnetic fields at very low temperature \cite{Dalfovo99}.
On the other hand, the appearance of the Coulomb potential barrier plays 
 an important role in confining the $\alpha$ bosons in the inside region.
It is noted that the barrier comes out naturally from the self-consistent calculation
 of the Gross-Pitaevskii equation.
The single $\alpha$ particle energy for the $3\alpha$ and $4\alpha$ systems are 
 smaller than the Coulomb potential barrier.
This means that the dilute states are quasi-stable against $\alpha$ decay.

Increasing the number of $\alpha$ bosons beyond $N=5$, the following 
 interesting features can be seen in the single $\alpha$ particle potentials:
 1)~the depth of the flat potential becomes shallower, and its range is 
 expanding to the outer region faster than $\propto N^{1/3}$, 
 and 2)~the height of the Coulomb potential barrier 
 is getting lower and almost disappears at around $N=10$. 
The first behavior of the potentials means that the $N\alpha$ system is inflating 
 with increase of $N$, as inferred from the behavior of the rms radius
 shown in Fig.~\ref{fig:2}.
The reason of why the depth of the potential becomes shallower is related to 
 the fact that the single $\alpha$ particle energy is getting larger.
It is given as follows:
From Table~\ref{tab:1}, the contribution from the kinetic energy to the single $\alpha$ 
 particle energy becomes smaller with increase of $N$, reflecting the inflation of 
 the dilute $N\alpha$ system.
This fact indicates that the Thomas-Fermi approximation is getting better, and then,
 the single $\alpha$ particle energy is given approximately as the contribution
 from only the potential energies. 
The potential energies consist of three contributions, namely, the two-range
 Gaussian term, density-dependent term, and Coulomb potential [see Eq.~(\ref{DD_pot})], 
 where the first and second ones are short-range, and the third is long-range.
When the dilute $N\alpha$ system is inflating with $N$ and the distance between 
 two $\alpha$ bosons is getting larger, the contribution from the long-range potential 
 should overcome steadily the one from the short-range one. 
Consequently, the potential depth (single $\alpha$ particle energy) becomes 
 gradually shallower (larger) and the height of the Coulomb barrier is getting lower,
 with increase of $N$.
The increase of the single $\alpha$ particle energy means that the total energy
 of the dilute $N\alpha$ system becomes larger with $N$.
The present results are consistent with the behavior
 of the $N$ dependence of the total energy (see Fig.~\ref{fig:2}).

The second interesting behavior of the single $\alpha$ particle potential 
 with $N\geq 5$ is that there exists a critical number of $\alpha$ 
 bosons, $N_{cr}$, beyond which the system is not confined
 anymore, as mentioned in Sec.~I.
Around $N=10$,  the Coulomb barrier has almost disappeared, and 
 the single $\alpha$ particle potential is nearly flat up to the outer region, 
 although the most outer region, dominated by only the Coulomb potential, 
 is falling to zero (not illustrated in Fig.~\ref{fig:3}).
Thus, we can roughly estimate the critical number as $N_{cr}\sim10$,
 namely, $^{40}$Ca. 

\begin{figure}
\begin{center}
\includegraphics*[scale=0.4]{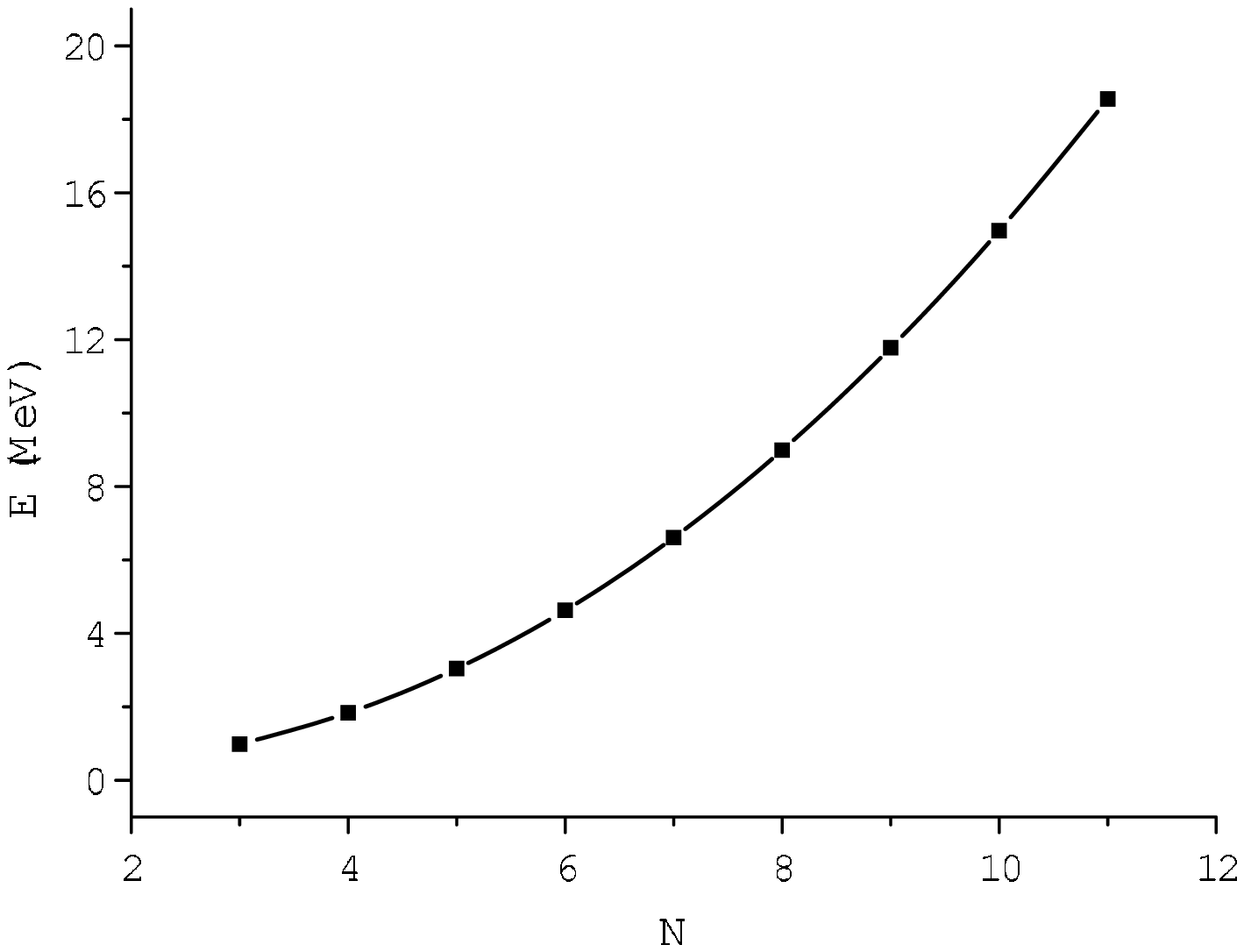}
\includegraphics*[scale=0.4]{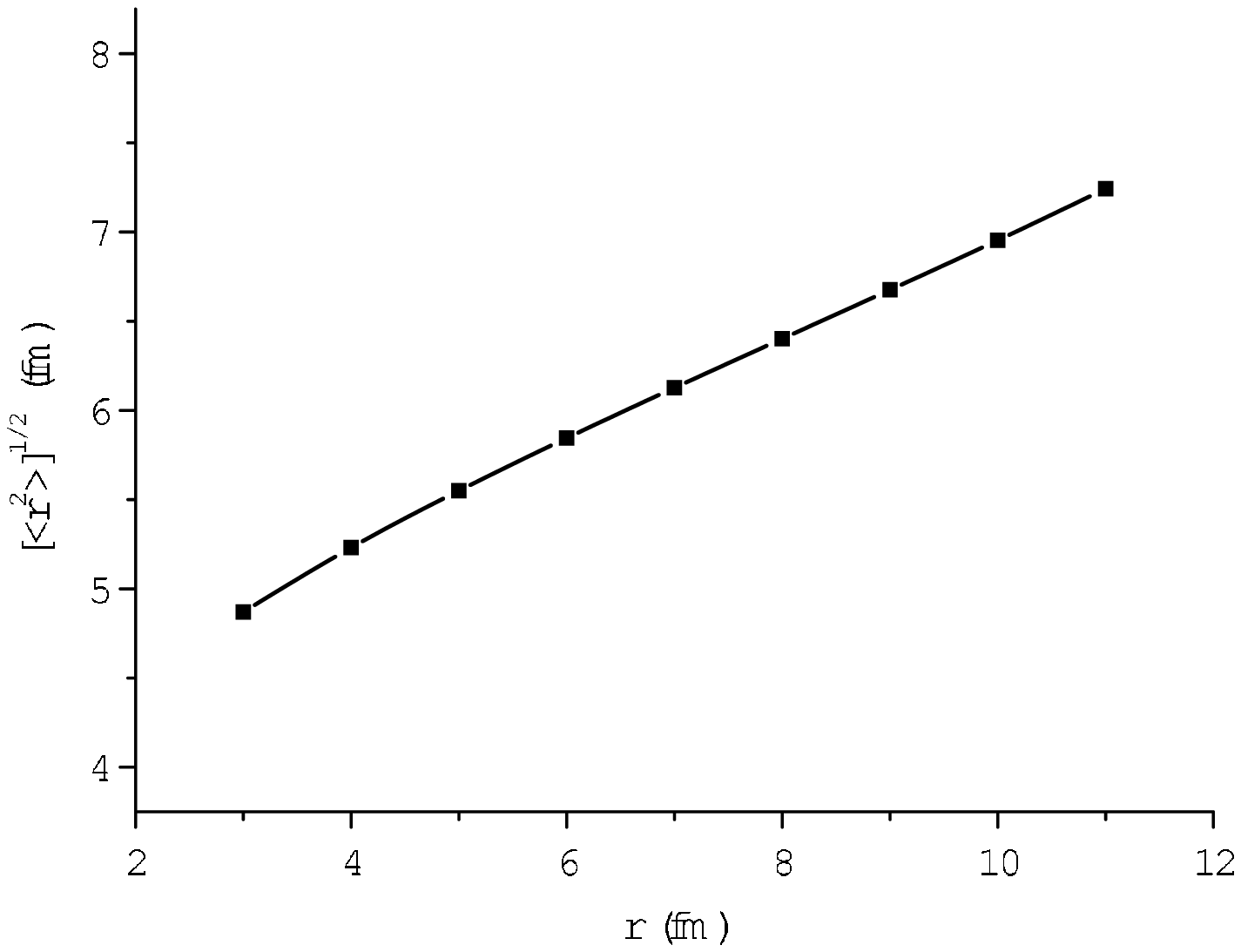}
\caption{
(a)~Total energies for the dilute $N\alpha$ states measured from each $N\alpha$
threshold, and (b)~their nuclear rms radii, which are obtained by solving 
the Gross-Pitaevskii equation with the phenomenological $2\alpha$ and $3\alpha$
potentials. 
}
\label{fig:4}
\end{center}
\end{figure}

It is interesting to compare the above results with those 
 obtained by using the Gross-Pitaevskii approach with 
 the phenomenological $2\alpha$ and $3\alpha$ potentials 
 given in Eqs.~(\ref{2_body}) and (\ref{3_body}), respectively.
The calculated total energies and rms radii for $N\alpha$ systems are shown
 in Fig.~\ref{fig:4}.
We find that they are in good agreement with those of the case with
 the density-dependent potential in the Gross-Pitaevskii equation
 within about 10\% except for the $N=3$ and $4$ cases.  
Table~\ref{tab:2} shows the single $\alpha$ particle energies and contributions
 from the kinetic energy and potential energies. 
Comparing them with those in Table~\ref{tab:1}, the $N$ dependence 
 of the kinetic energy, Coulomb potential energy and the sum of other potential 
 energies is rather similar to the one with the density-dependent potential 
 in the Gross-Pitaevskii equation except for the $N=3$ and $4$ systems.
This is surprising if taking into account the fact that the forms of the two kinds of
 the effective $\alpha$-$\alpha$ potentials are quite different from one to another. 
These results might indicate that the dilute $N\alpha$ systems with $N\geq 5$ 
 do not depend strongly on the details of the effective $\alpha$-$\alpha$ potential, 
 while those with a small-number-$N\alpha$ ($N=3$ and 4) are
 sensitive to the potential.

\begin{figure}
\begin{center}
\begin{minipage}{0.90\linewidth}
\includegraphics*[width=\linewidth]{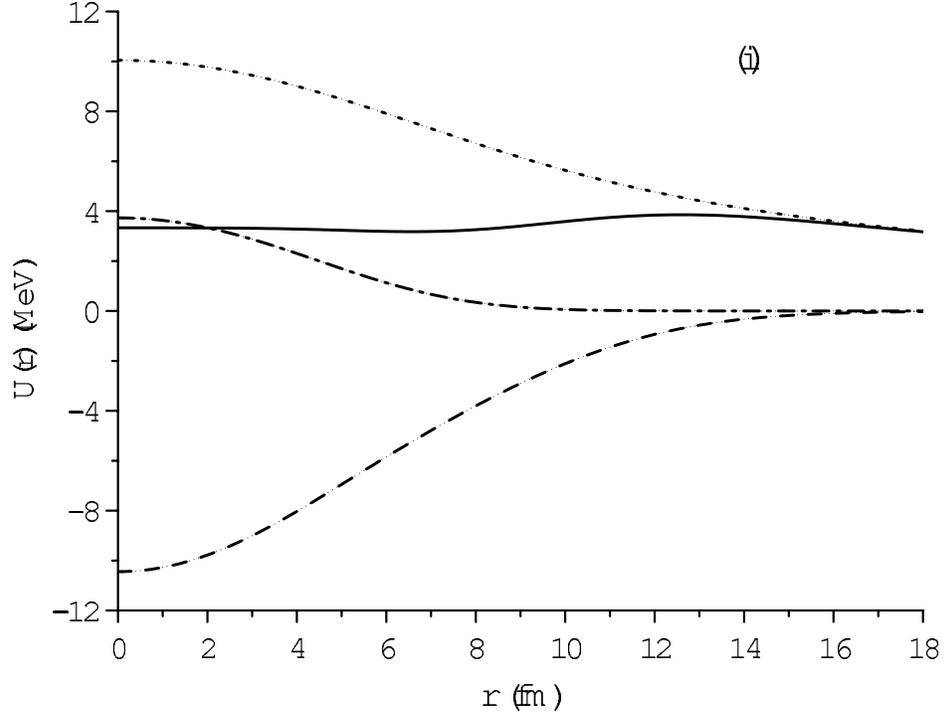}
\end{minipage}
\caption{
Single $\alpha$ particle potentials $U_\alpha(R)$ (solid line) 
which are obtained by solving 
the Gross-Pitaevskii equation with the phenomenological
$2\alpha$ and $3\alpha$ potentials;
(a)~$3\alpha$, (b)~$4\alpha$, (c)~$5\alpha$, (d)~$6\alpha$,
(e)~$7\alpha$, (f)~$8\alpha$, (g)~$9\alpha$, (h)~$10\alpha$, 
(i)~$11\alpha$ systems.
The dashed, dot-dashed and dotted lines demonstrate, respectively, the contribution 
from the $2\alpha$ potential, $3\alpha$ potential and Coulomb potential. 
}
\label{fig:5}
\end{center}
\end{figure}

The single $\alpha$ particle potentials are shown in Fig.~\ref{fig:5}.
For the $3\alpha$ and $4\alpha$ systems, we see the almost flat potential behavior 
 in the inside region, and the Coulomb potential barrier in the outer region, while
 increasing the number of $\alpha$ bosons from $N=5$, the depth of the flat potential 
 becomes shallower, and its range is expanding to the outer region, and the height of 
 the Coulomb potential barrier is getting lower and almost disappears at around $N=10$. 
The qualitative potential behaviors are almost the same as those in case of 
 the density-dependent potential with the Gross-Pitaevskii equation (see Fig.~\ref{fig:3}).
In fact, the behavior of the single $\alpha$ particle potential in Fig.~\ref{fig:5}
 is in good agreement with that in Fig.~\ref{fig:3} within about 10~\%.
On the other hand, we can conjecture the critical number $N_{cr}$ from 
 the behavior of the single $\alpha$ potentials in Fig.~\ref{fig:5},
 which is estimated roughly as $N_{cr}\sim10$, the result being
 the same as that of the case with the density-dependent potential.

\subsection{$N\alpha$ systems in the Hill-Wheeler-equation approach}

The Gross-Pitaevskii equation is simple and useful for the study of the structure of 
 the dilute $N\alpha$ system, as seen in the previous subsection.
The center-of-mass motion in the $N\alpha$ system, however, is not completely 
 removed in this framework.
The Hill-Wheeler-equation approach is free from the center-of-mass motion 
 of the $N\alpha$ system. 
The effect of the center-of-mass motion is expected to be non-negligible
 for the total energy and rms radius etc.~in small-number $\alpha$-boson 
 systems, in particular, $N=3$ and 4.
Thus, it is important to study the dilute $N\alpha$ systems with the use of 
 the Hill-Wheeler equation.
The two approaches should give the same results for dilute $N\alpha$ systems
 with rather large number of $N$.
In this subsection, we first demonstrate that the Hill-Wheeler-equation approach is useful 
 to describe the dilute $N\alpha$ states with the spherical shape, and the results 
 are briefly compared with those from the Gross-Pitaevskii equation.
Then, the approach is applied to the deformed $N\alpha$ systems ($J^\pi=0^+$) 
 with axial symmetry.
We use the phenomenological $2\alpha$ and $3\alpha$ potentials of Eqs.~(\ref{2_body}) 
 and (\ref{3_body}) as the effective $\alpha\alpha$ potential.

\subsubsection{Spherical $N\alpha$ systems}

\begin{figure}
\begin{center}
\includegraphics*[scale=0.4]{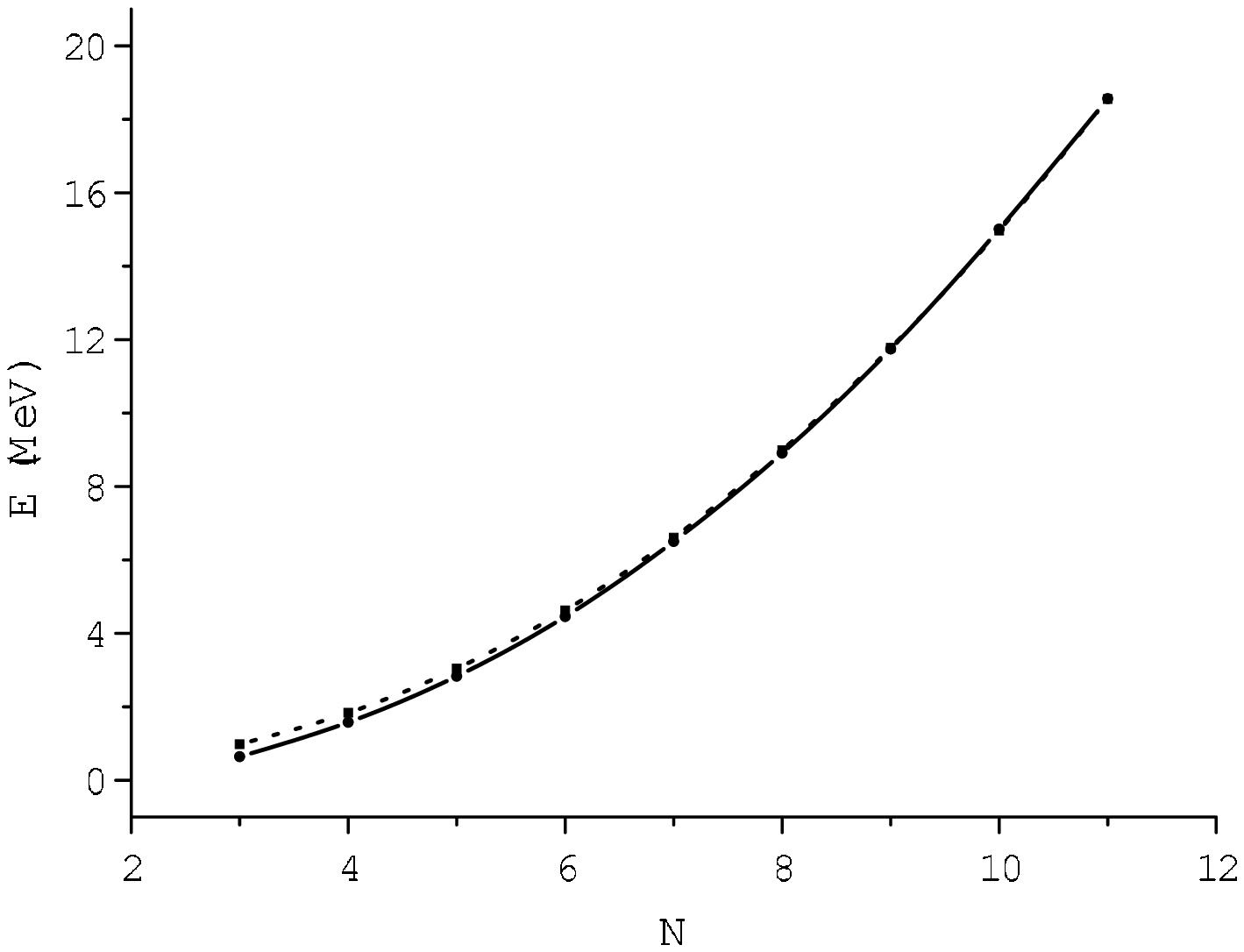}
\includegraphics*[scale=0.4]{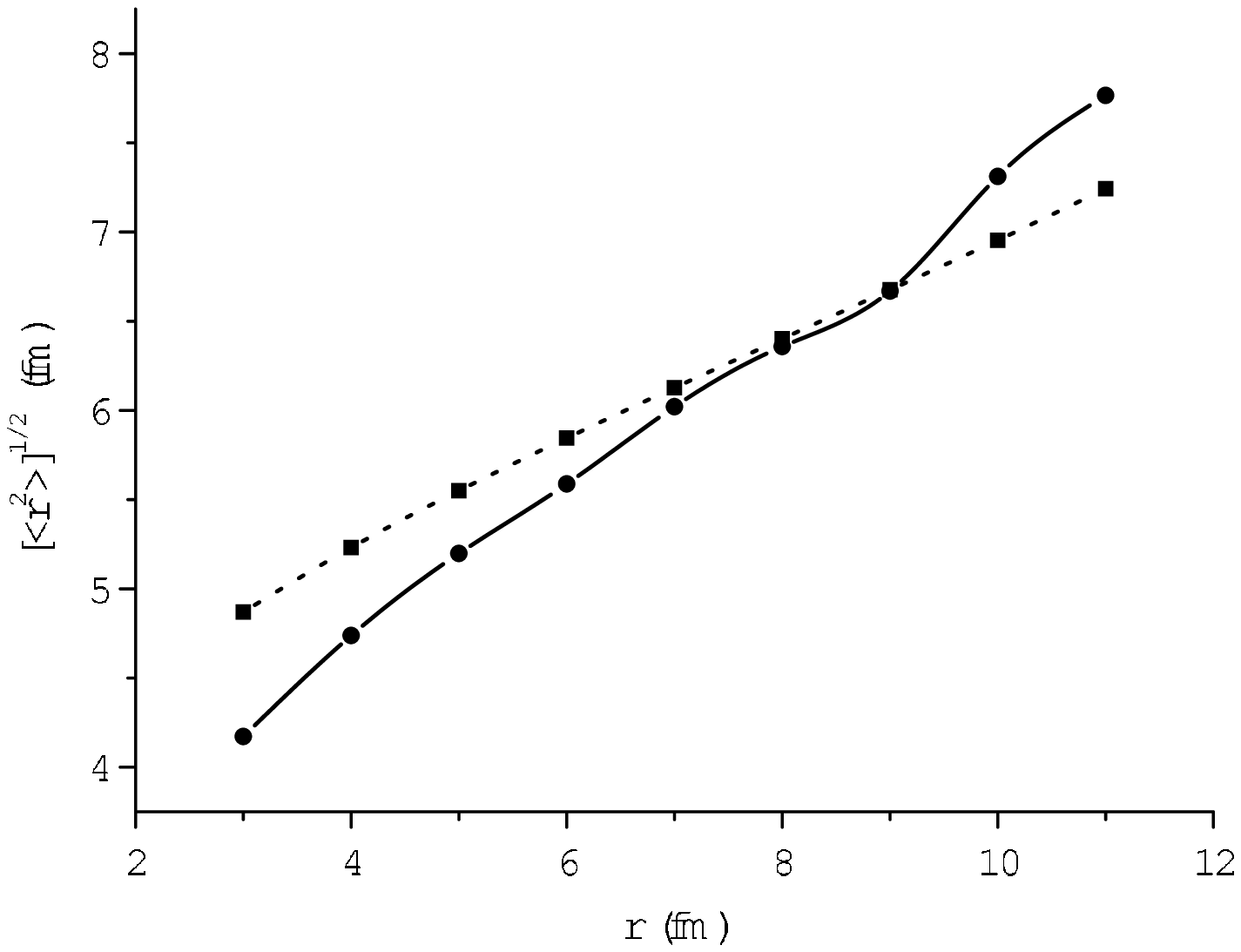}
\caption{
(a)~Total energies for the dilute $N\alpha$ states (spherical case) measured from each $N\alpha$
threshold (solid line), and (b)~their nuclear rms radii (solid), which are obtained by solving 
the Hill-Wheeler equation with the phenomenological $2\alpha$ and $3\alpha$
potentials. For comparison, we also give the corresponding values obtained by solving 
the Gross-Pitaevskii equation 
with use of the same potentials (dotted lines). 
}
\label{fig:6}
\end{center}
\end{figure}

The calculated total energies and rms radii for spherical $N\alpha$ states
 are illustrated in Fig.~\ref{fig:6}, in which we show those obtained by
 solving the Gross-Pitaevskii equation with the same effective 
 $\alpha$-$\alpha$ potential for comparison.
We found that the calculated total energies in the Hill-Wheeler approach
 are almost the same as those in the Gross-Pitaevskii approach.
This indicates that the center-of-mass kinetic energy correction in Eq.~(\ref{GPE})
 is a very good approximation to remove the effect in the total energy.
The non-negligible deviation between the two frameworks, however, 
 can be seen in the rms radii, in particular, we have about 20~\% deviation 
 for the $3\alpha$ system, although that is getting smaller with increasing $N$ and 
 it is almost zero in the $7\alpha\sim9\alpha$ systems 
 (the discrepancy in the $10\alpha$ and $11\alpha$ systems will be discussed later).
The results tell us that the center-of-mass correction for the rms radius 
 in the Gross-Pitaevskii-equation approach [see Eqs.~(\ref{rms_N}) and (\ref{rms_N_cor})] 
 is not very good for the $3\alpha$ and $4\alpha$ systems, while it is relatively good 
 for the $N\alpha\geq 5\alpha$ systems, 
 and the difference between the two approaches is less than about 10~\%, 
 and it is diminishing with increasing $N$.

In Fig.~\ref{fig:6} we see that the deviation of the rms radius appears again
 in the $10\alpha$ and $11\alpha$ systems.
The reason of why this deviation occurs can be understood, recalling the 
 following facts:
According to the results of the Gross-Pitaevskii equation
 with the phenomenological $2\alpha$ and $3\alpha$ potentials, 
 the $10\alpha$ system is the critical one which can not exist as a nuclear 
 state, as mentioned in the previous section.
In such a critical system, the behavior of the wave functions in the outer (inner) 
 region is very sensitive (not very sensitive) to how to solve the equations and 
 to obtain the wavefunctions.
The value of the rms radius (total energy) is generally sensitive (not very sensitive) 
 to the behavior of the wave function in the outer region.
Thus, the re-occurrence of the deviation in the rms radii indicates
 that the $10\alpha$ system is critical in the Hill-Wheeler-equation approach.
 
The above results show us that the Hill-Wheeler-equation approach is very useful
 to describe the dilute $N\alpha$ states as well as the Gross-Pitaevskii-equation approach.
Thus, we can apply the approach to the deformed $N\alpha$ system with the axial deformation.

\subsubsection{Deformed $N\alpha$ systems with $J^\pi=0^+$}

\begin{figure}
\begin{center}
\includegraphics*[scale=0.4]{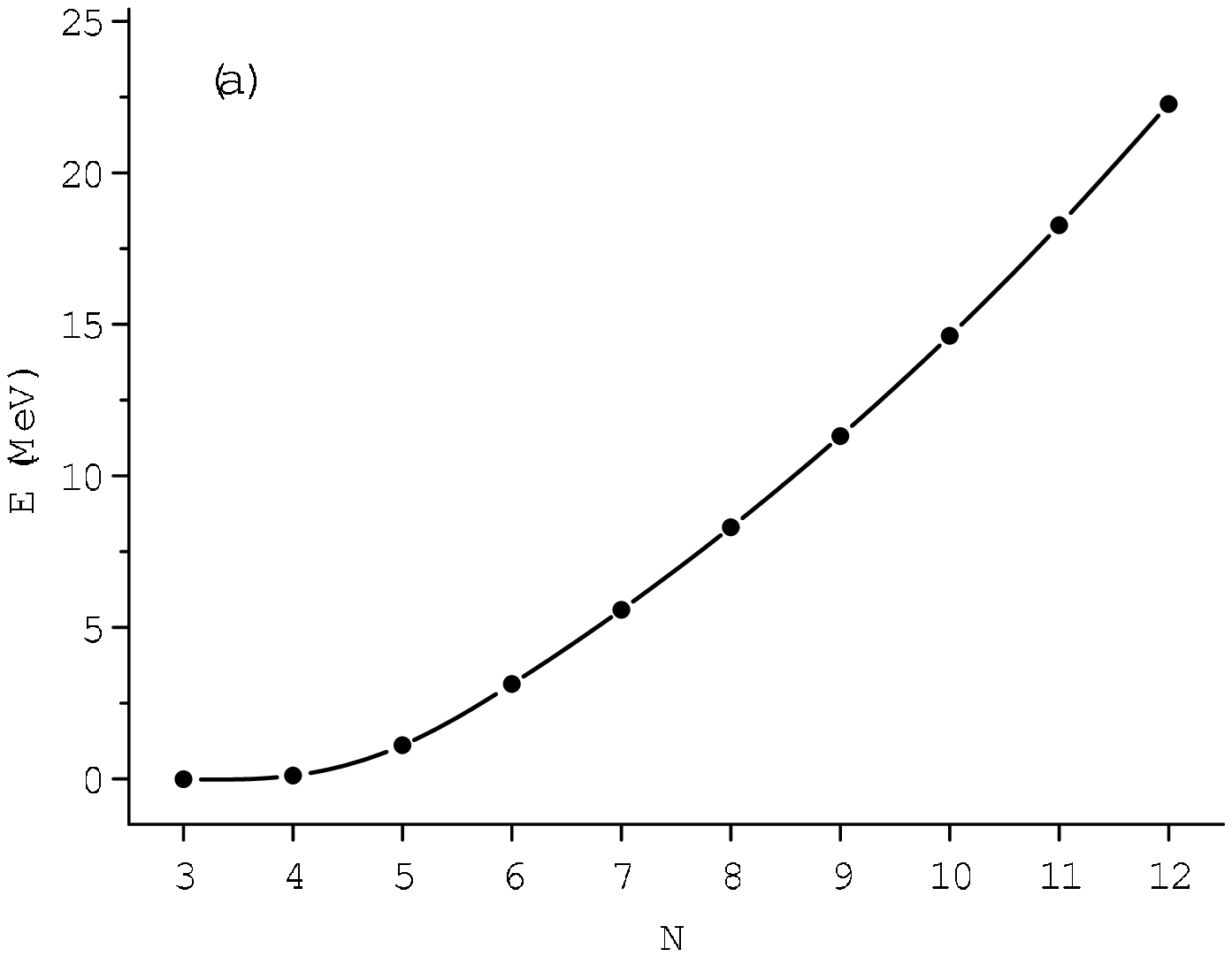}
\includegraphics*[scale=0.4]{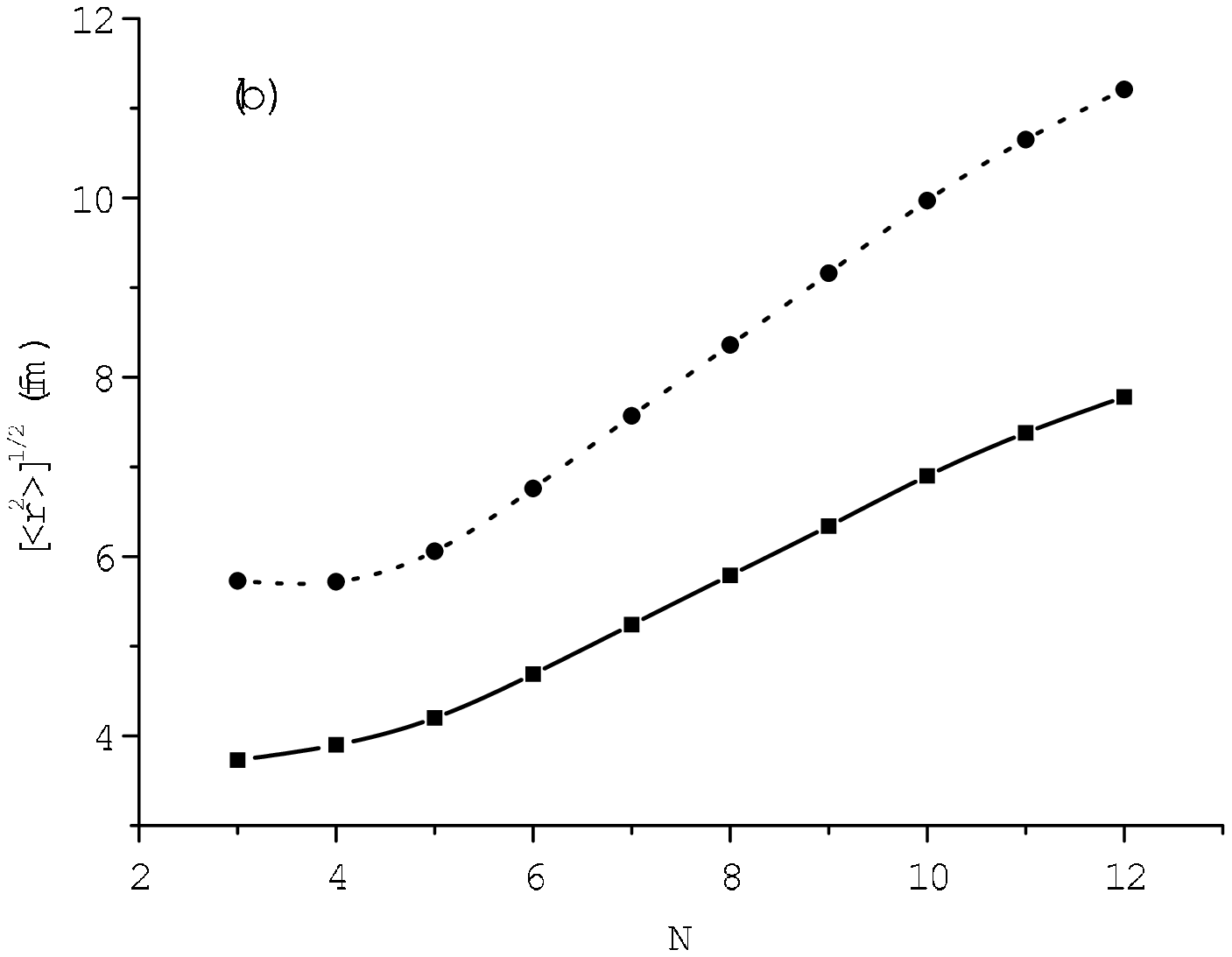}
\caption{
(a)~Total energies for the dilute $N\alpha$ states with the axial deformation 
measured from each $N\alpha$ threshold (solid line), and (b)~their nuclear rms radii 
(solid) and rms distance between $2\alpha$ particles (dotted), 
which are obtained by solving the Hill-Wheeler equation with 
the phenomenological $2\alpha$ and $3\alpha$ potentials.  
}
\label{fig:7}
\end{center}
\end{figure}

The $N\alpha$ states ($J^\pi=0^+$) with the axial deformation are obtained 
 by solving the Hill-Wheeler equation in Eq.~(\ref{Hill_Wheeler}) with the Gaussian size 
 parameter set A given in Table~\ref{tab:3}.
Figure~\ref{fig:7} illustrates the calculated energies, nuclear rms radii defined 
 in Eq.~(\ref{rms_N}) and the rms distances between $2\alpha$ bosons.
The detailed values are shown in Table~\ref{tab:4}.
The rms distance between $2\alpha$ bosons ($6\sim11$ fm) for $N=3\sim12$ 
 is considerably larger than that of the $^8$Be ground state ($\sim4$ fm).
The result indicates that the $N\alpha$ states obtained here are of
 very dilute $N\alpha$ structure.

The total energy of the $N\alpha$ state increases gradually with $N$, 
 although those for $N$=3 and 4 are not changed very much.
The latter is in contrast to the results of the spherical case together
 with those for the Gross-Pitaevskii-equation approach (see Figs.~\ref{fig:2}, 
 \ref{fig:4} and \ref{fig:6}).
In the $3\alpha$ and $4\alpha$ systems, the total energies and nuclear rms radii 
 are respectively given as follows; $E=-0.01$ and 0.13 MeV, 
 and $\sqrt{\langle r^2_N\rangle}$=3.73 and 3.90 fm.
The values are in good correspondence to those for the condensate $\alpha$ cluster 
 states discussed by Tohsaki et al.~\cite{Tohsaki01}, where their calculated results are 
 $E^{cal}$=0.5 and $-0.7$ MeV (vs. $E^{exp}$=0.38 and $-0.44$ MeV) and 
 $\sqrt{\langle r^2_N\rangle}$=4.29 and 3.97 fm, 
 respectively, for the dilute $3\alpha$ and $4\alpha$ states.  
It is instructive here to compare the present results with those in the spherical 
 $N\alpha$ states in order to estimate the effect of deformation.
The comparison of the total energy and rms radius is given in Table~\ref{tab:4}.
The energy gains (reductions of rms radius) due to the deformation 
 in the $3\alpha$, $4\alpha$ and $5\alpha$ systems are, respectively, 
 0.6, 1.5 and 1.7 MeV  (11~\%, 18~\% and 19~\%), while this is getting smaller 
 for $N\geq 6$.
Thus, we found that the deformation effect is significant for 
 relatively small-number $\alpha$ systems, and shows a good correspondence
 of our results for the $3\alpha$ and $4\alpha$ states with those 
 by Tohsaki et al.~\cite{Tohsaki01}. 
The present result that the dilute $N\alpha$ states ($J^\pi=0^+$) may be deformed 
 is natural if taking into account the fact that a gas-like $N\alpha$ state with
 relatively small number can easily be deformed.
If it is right, there may exist dilute $N\alpha$ nuclear states with $J^\pi=2^+$ and 
 $4^+$ etc.
In fact, a candidate of the dilute $3\alpha$ state with $J^\pi=2^+$ is observed
 at $E$=3.3 MeV measured from the $3\alpha$ threshold \cite{Ajzenberg90}.
The dilute multi $\alpha$ cluster states with non-zero angular momentum
 will be discussed elsewhere.   

\begin{figure}
\begin{center}
\includegraphics*[scale=0.4]{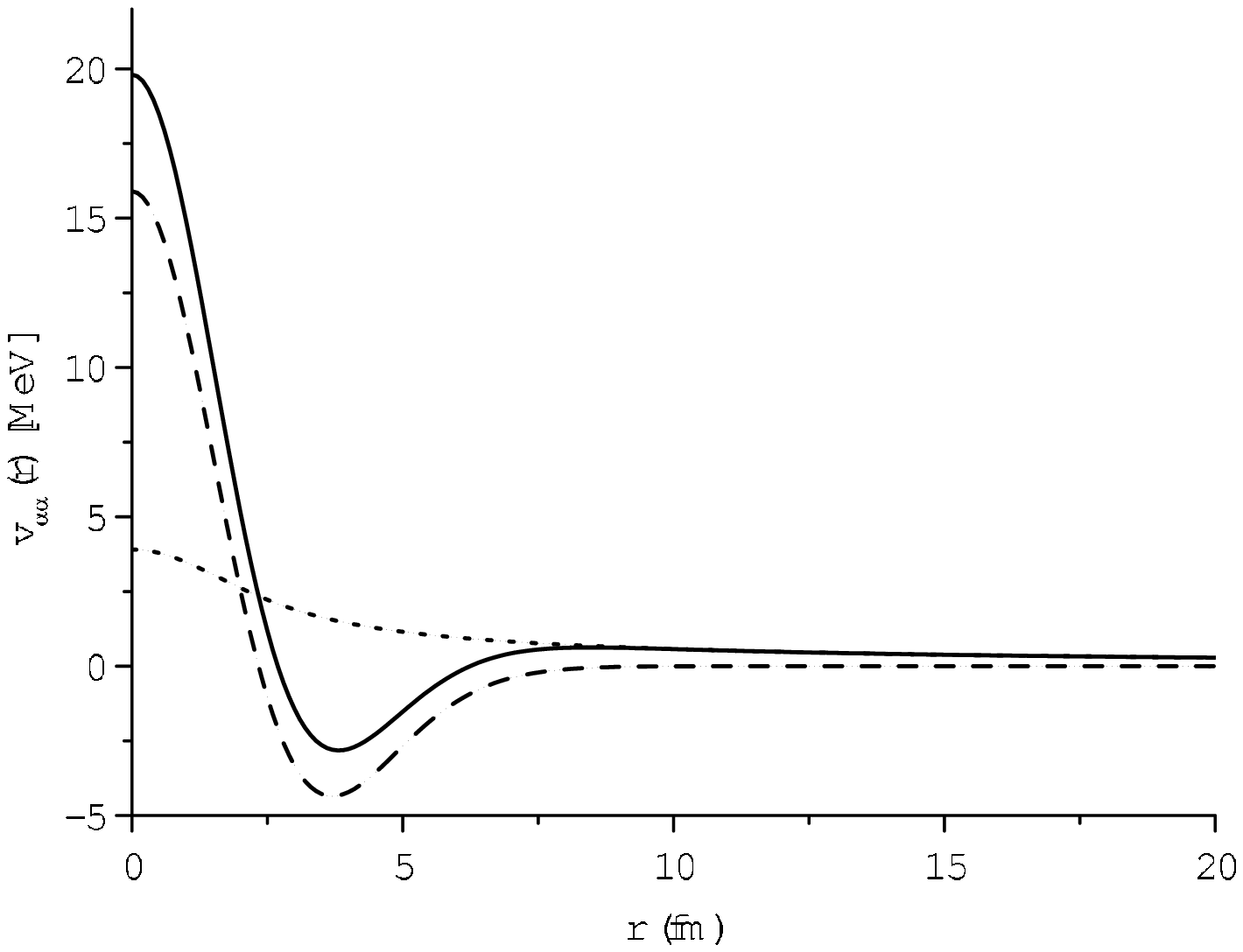}
\caption{
Phenomenological $2\alpha$ potential (solid line), where the $2\alpha$
nuclear potential [Eq.~(\ref{2_body})] and its Coulomb potential [Eq.~(\ref{Coulomb_pot})] 
are drawn by the dashed and dotted lines, respectively. 
}
\label{fig:8}
\end{center}
\end{figure}

From Fig.~\ref{fig:7}(a), we notice that the gas-like $N\alpha$ states 
 with $N\geq5$ appear above the $N\alpha$ threshold, not close to it, 
 in contrast to the fact that the dilute $3\alpha$ and $4\alpha$ states are 
 located in the vicinity of their respective thresholds.
This is a non-trivial result of the present calculation.
The reason of why the energy of the dilute $N\alpha$ states increases
 with $N$ is given as follows.
Let us consider the effective energy of the $"2\alpha"$ system 
 in the dilute $N\alpha$ state, which corresponds to the quantity of the 
 total energy of the $N\alpha$ state divided by the number of $\alpha$ pairs
 which is given in Table~\ref{tab:4}.
Since the dilute character allows us to neglect approximately the kinetic 
 energy, the effective energy of the $"2\alpha"$ system could be given mainly 
 as the sum of the $2\alpha$ nuclear potential energy and its Coulomb
 potential energy (the contribution from the $3\alpha$ potential may be
 neglected approximately because of the very small amount).
Figure~\ref{fig:8} shows the $2\alpha$ potential used in the
 present study, where we also show the $2\alpha$ nuclear potential 
 [see Eq.~(\ref{2_body})] and its Coulomb potential [see Eq.~(\ref{Coulomb_pot})].
The attraction of the former is largest around 4 fm and is negligible 
 beyond about 7 fm, while the Coulomb potential is long-ranged and 
 its repulsion is substantial even beyond 7 fm.
Increasing the number of the $\alpha$ particles, the gas-like 
 $N\alpha$ system is expanding and the rms distance between 
 two $\alpha$ particles is also becoming larger ($6\sim10$ fm) in the case
 of $N\geq5$, as shown in Fig.~\ref{fig:7}.
Then, the attractive contribution from the $2\alpha$ nuclear potential 
 to the effective $"2\alpha"$ energy should become noticeably smaller with 
 increasing $N$, reflecting the short-range (attractive) behavior of the potential.
The repulsive contribution from the Coulomb potential, however, should not be
 much smaller than in the case of the $2\alpha$ nuclear potential, and dominate
 over the contribution from the nuclear potential, because of the long-range 
 (repulsive) behavior.
These facts explain the raise of the effective $"2\alpha"$ energy with $N$, 
 which means that the energy of the dilute $N\alpha$ states increases
 with $N$ and does not stay around the threshold energy.   
In Table~\ref{tab:4} we can find quantitatively the increase of the effective 
 $2\alpha$ system with $N$.
The gradual dominance of the total Coulomb potential energy over the total
 $2\alpha$ nuclear potential energy in the $N\alpha$ system can be also seen 
 in Table~\ref{tab:4}.

It is interesting to see the role of the $3\alpha$ potentials in the dilute $N\alpha$ system. 
From Table~\ref{tab:4}, the contribution to the total energy is small
 in the case of $^{12}$C and $^{16}$O, while it gives a non-negligible effect 
 with increasing $N$, as seen in Table~\ref{tab:3}, 
 because it is proportional to the number of $N(N-1)(N-2)/6$.
In fact, the energy $\langle V_3\rangle$ becomes steadily larger 
 with the number of $\alpha$ bosons, although the quantity 
 of the energy divided by the number of trios of the $\alpha$ bosons 
 is decreasing with $N$, reflecting the fact that the $N\alpha$ 
 system is expanding with $N$.
It is instructive to study the case of no $3\alpha$ potentials within
 the present framework.
The calculated results show that the $N\alpha$ system gradually falls into
 a collapsed state with increasing $N$ as discussed before; 
 for example, $E\sim-55$ MeV and $\sqrt{\langle r^2_N\rangle}=4.77$ fm
 for $N=40$.
The result tells us that the dilute $\alpha$ boson states appear
 under the cooperative work between the two-body and three-body 
 potentials in the present model.

Finally, we estimate the critical number $N_{cr}$ of the dilute $N\alpha$
 state beyond which the system is unbound.
Since the critical state should be very unstable, it is difficult to determine it
 exactly with the present Hill-Wheeler-equation approach.
However, we can deduce it approximately from studying the stability 
 of the calculated eigen energies and rms radii against changing 
 the model space determined by the Gaussian size parameters 
 in Eqs.~(\ref{gs_para_1}) and (\ref{gs_para_3}).
The procedure is as follows.
The present framework is able to describe dilute states 
 trapped by the Coulomb barrier under the condition that we choose the 
 range of the Gaussian size parameters wide enough to cover a configuration 
 space over the whole rms radius of the states.
If the state is stable or relatively stable, the calculated energies and rms 
 radii are not changed very much against the variation of 
 the Gaussian size parameters. 
Otherwise, those depend sensitively on the choice of the parameters. 
Using the wide model space, however, a special attention must 
 be paid to investigate the eigen-states obtained from the Hill-Wheeler equation,
 because a discretized continuum state, which is unphysical,
 has a chance to become the lowest state in energy in the present
 variational calculation.
We can easily identify it by investigating the behavior of its energy and/or
 rms radius against changing the Gaussian size parameters.
The unphysical state usually has an abnormally long rms
 radius, which is a little smaller than or almost the same as 
 the value of $b_1^{(k_{max})}$ or $b_3^{(K_{max})}$ in Eqs.~(\ref{gs_para_1})
 and (\ref{gs_para_3}).   
Table~\ref{tab:5} shows the calculated results for the two parameter 
 sets A and B (see Table~\ref{tab:3}).
The calculated energies and rms radii for the dilute $N\alpha$ states 
 with $N=3\sim10$ are almost the same  for both the parameter 
 sets A and B, while we see some discrepancies in those with $N\geq 11$,
 in particular, in the rms radii, for the two parameter sets.
Even if we use other Gaussian size parameter sets which cover 
 the configuration space over the rms radius for the dilute states,
 the results obtained are found to be similar to those mentioned above.
Thus, the critical number of the dilute $N\alpha$ boson systems is 
 estimated roughly to be $N_{cr}\sim10$, namely, $^{40}$Ca, in the present study.
It is noted that the number is the same as that in the spherical case
 as well as in the Gross-Pitaevskii-equation approach. 

\section{Summary}
%
The dilute $N\alpha$ cluster condensate states with $J^\pi=0^+$ have been studied with 
 the Gross-Pitaevskii equation (GP) and the Hill-Wheeler equation (HW), 
 where the $\alpha$ cluster is treated as a structureless boson.
Two kinds of effective $\alpha$-$\alpha$ potentials were used:
 the density-dependent $\alpha$-$\alpha$ potential (DD) and the phenomenological 
 $2\alpha$ potential plus 3-body $\alpha$ potential (PP), and we also included
 the folded Coulomb potential between the $2\alpha$ bosons.
Both potentials (only the latter) were (was) applied to study the structure of 
 the spherical (spherical and axially deformed) dilute $N\alpha$ states 
 with the Gross-Pitaevskii equation (Hill-Wheeler equation).
Thus, we have studied the dilute multi $\alpha$ cluster states with the four
 different frameworks; (equation, potential, shape)=(GP, DD, S), (GP, PP, S),
 (HW, PP, S) and (HW, PP, D), where S and D denotes the spherical
 and deformed shapes, respectively.

The main results to be emphasized here are as follows:

1)~All of the $N\alpha$ states obtained show the dilute $N\alpha$ structure.
They are common to all of the four different frameworks.
The reason of why the total energy of the gas-like $N\alpha$ state increases gradually
 with $N$ and do not remain around their $N\alpha$ threshold values is understood 
 as the competition between the nuclear $\alpha$-$\alpha$ nuclear potential (attractive)
 and its Coulomb potential (repulsive).
In fact, increasing the number of $N$, the $\alpha$-$\alpha$ distance is becoming larger
 ($6\sim12$ fm for $N=3\sim12$), and then the contribution from the $\alpha$-$\alpha$ nuclear 
 potential per $2\alpha$ pair in the dilute $N\alpha$ state is decreasing rapidly 
 because of the short-range character, while that from 
 the Coulomb one is decreasing very slowly and remains almost constant
 for $N=5\sim12$, reflecting the $1/r$ character of the Coulomb potential.  

2)~The $N$ dependence of the behavior of the calculated single $\alpha$ potentials
 obtained by solving the Gross-Pitaevskii equation within the (GS, DD, S) and
 (GW, PP, S) frameworks is impressive.
For the $3\alpha$ and $4\alpha$ systems, we see an almost flat potential behavior 
 in the inside region, and the Coulomb potential barrier in the outer region, while
 increasing the number of $\alpha$ bosons from $N=5$, the depth of the flat potential 
 becomes shallower, and its range is expanding to the outer region, while the height of 
 the Coulomb potential barrier is decreasing and almost disappears at around $N=10$. 
The origin of the appearance of the flat region is mainly due to the validity of the
 Thomas-Fermi approximation, namely, the neglect of the contribution from 
 kinetic energy because of the dilute character of the $N\alpha$ system.
The result is analogous to the dilute atomic condensate state trapped by the magnetic
 fields at very low temperature.
On the other hand, the results for the total energy and rms radius were 
 quite similar for the above-mentioned two frameworks except for $N$=3 and 4.
The result might indicate that the dilute $N\alpha$ states with $N\geq 5$ are not
 very sensitive to the details of the effective $\alpha$-$\alpha$ potential (DD or PP).  

3)~Comparing the results from the Gross-Pitaevskii equation (GP)  
 and Hill-Wheeler equation (HW), we could see the effect of the neglect of 
 the center-of-mass motion in the Gross-Pitaevskii equation and also the usefulness 
 of the Hill-Wheeler-equation approach 
 for describing the dilute $N\alpha$ system.
In the small-number $N\alpha$ systems, the effect was found to be non-negligible, 
 in particular, for the rms radius, while there is less than about 10~\% deviation 
 for the $N\alpha\geq 5\alpha$ systems.

4)~The axial deformation effect in the dilute $N\alpha$ systems is substantial 
for the small-number $\alpha$ systems, but it is getting smaller for $N\geq 6$.
In fact, the energy gain and reduction of the rms radius due to the deformation
 are $0.6\sim1.7$ MeV and $11\sim19$ \%, respectively, for the former system.
Due to the effect, the calculated results of the total energy and rms radius 
 for $N$=3 and 4 in the (HW, PP, D) framework are improved from
 those in the (HW, PP, S) framework, and give a good correspondence with
 those for the condensate $\alpha$ cluster states discussed by Tohsaki 
 et al.~\cite{Tohsaki01}
The present result that the dilute $N\alpha$ states ($J^\pi=0^+$) may be deformed 
 is natural, taking into account the fact that a gas-like $N\alpha$ state with
 relatively small number of $\alpha$'s could easily be deformed.

5)~We estimated the critical number of the $\alpha$ bosons, $N_{cr}$, beyond which 
 the system is unbound for the four frameworks,
 (GP, DD, S), (GP, PP, S), (HW, PP, S) and (HW, PP, D).
All of the frameworks indicated that the number is roughly $N_{cr}\sim10$, which
 is not strongly dependent on the frameworks.
Thus, we concluded that the dilute $N\alpha$ cluster states could exist in
 the $^{12}$C to $^{40}$Ca systems with $J^\pi=0^+$, whose energies vary from 
 threshold up to about 20 MeV in the present calculation.

The estimate of $N_{cr}\sim10$ is of course subject to the validity of our phenomenological
 approach treating the $\alpha$-particles as ideal bosons.
We, however, believe, for reasons outlined in the paper, that our estimate for $N_{cr}$ may
 be correct to within $\pm20$~\%.
In any case the value for $N_{cr}$, {\it i.e.} the maximum of $\alpha$-particles in the
 condensate state is relatively modest.
A very interesting question in this context is whether adding a few neutrons may stabilize
 the condensate and thus allow for much higher numbers of condensed $\alpha$'s.
One should remember that $^8$Be is (slightly) unbound whereas $^9$Be and $^{10}$Be 
 are bound.

Concerning experimental detection of the $\alpha$-condensates, 
 the decay scheme of the dilute $N\alpha$ state is conjectured to proceed mainly 
 via $\alpha$ decay.
This indicates that such systems may be observed through 
 the following sequential $\alpha$ decays:
 {[dilute $N\alpha$ state]}~$\rightarrow$~{[dilute $(N-1)\alpha$ state]}+$\alpha$,
 {[dilute $(N-1)\alpha$ state]}~$\rightarrow$~{[dilute $(N-2)\alpha$ state]}+$\alpha$,
 $\cdots$. 
Therefore, the sequential $\alpha$ decay measurement is expected
 to be one of the promising
 tools to search for the dilute multi-$\alpha$ cluster states, produced via
 the $\alpha$ inelastic reaction, heavy-ion collision reaction, and so on.  
It is highly hoped to perform such the experiments in near future.

\section*{Acknowledgments}
We greatly acknowledge helpful discussions with Y.~Funaki, H.~Horiuchi, K.~Ikeda, 
 G.~R\"opke, and A.~Tohsaki.
One of the authors (T.Y.) is very grateful to the Theory Group of the Institut 
 de Physique Nucl\'eaire, Universit\'e Paris-Sud (Paris XI) 
 for its hospitality during his sabbatical year.

\clearpage
\begin{table}
\caption{
Calculated results of the Gross-Pitaevskii equation with the density-dependent $\alpha$-$\alpha$
 potential in Eq.~(\ref{DD_pot}); single $\alpha$ particle energy $\varepsilon$ and 
 contributions from the kinetic energy {$\langle t\rangle$}, two-range-Gaussian term {$\langle \upsilon_{2G}\rangle$},
 density-dependent term {$\langle \upsilon_D\rangle$} and
 Coulomb potential {$\langle \upsilon_C\rangle$} in Eq.~(\ref{DD_pot}).
The total energy and nuclear rms radius for the N$\alpha$ system are denoted as $E$ and 
 $\sqrt{\langle r^2_N\rangle}$, respectively.
All energies and rms radii are given, respectively, in units of MeV and fm.}
\label{tab:1}
\begin{center}
\begin{tabular}{ccccccccc}
\hline
\hline
\hspace{3mm}$N$\hspace{3mm} & \hspace{3mm}{nucleus}\hspace{3mm} 
     & \hspace{5mm}$\varepsilon$\hspace{5mm} & \hspace{5mm}$\langle t\rangle$\hspace{5mm}
     & \hspace{5mm}$\langle \upsilon_{2G}\rangle$\hspace{5mm} & \hspace{5mm}$\langle \upsilon_C\rangle$\hspace{5mm}  
     & \hspace{5mm}$\langle \upsilon_{D}\rangle$\hspace{5mm}  & \hspace{5mm}$E$\hspace{5mm} 
     & \hspace{5mm}$\sqrt{\langle r^2_N\rangle}$\hspace{5mm} \\  
\hline
 3 & $^{12}$C  & 0.18 & 0.38 & $-3.42$ & 2.31 & 0.91 & 0.38 & 4.14 \\
 4 & $^{16}$O  & 0.78 & 0.32 & $-3.74$ & 3.04 & 1.16 & 1.44 & 4.91 \\
 5 & $^{20}$Ne & 1.32 & 0.28 & $-3.90$ & 3.69 & 1.25 & 2.97 & 5.53 \\
 6 & $^{24}$Mg & 1.82 & 0.25 & $-3.95$ & 4.25 & 1.26 & 4.94 & 6.07 \\
 7 & $^{28}$Si  & 2.28 & 0.22 & $-3.91$ & 4.76 & 1.21 & 7.36 & 6.58 \\
 8 & $^{32}$S   & 2.72 & 0.20 & $-3.80$ & 5.20 & 1.20 & 10.2 & 7.05 \\
 9 & $^{36}$Ar & 3.13 & 0.19 & $-3.65$ & 5.59 & 1.00 & 13.4 & 7.51 \\
10 & $^{40}$Ca & 3.51 & 0.18 & $-3.45$ & 5.93 & 0.86 & 17.0 & 7.98 \\
11 & $^{44}$Ti & 3.87 & 0.18 & $-3.19$ & 6.19 & 0.69 & 21.0 & 8.46 \\
\hline
\hline
\end{tabular}
\end{center}
\end{table}

\clearpage
\begin{table}
\caption{
Calculated results of the Gross-Pitaevskii equation with the phenomenological $2\alpha$ and $3\alpha$
 potentials in Eqs.~(\ref{2_body}) and (\ref{3_body}); single $\alpha$ particle energy $\varepsilon$ and 
 contributions from the kinetic energy {$\langle t\rangle$}, $2\alpha$ potential {$\langle \upsilon_2\rangle$}
 in Eq.~(\ref{2_body}), $3\alpha$ potential {$\langle \upsilon_3\rangle$} in Eq.~(\ref{3_body}) and
 Coulomb potential {$\langle \upsilon_C\rangle$}.
The total energy and nuclear rms radius for the N$\alpha$ system are denoted as $E$ and 
 $\sqrt{\langle r^2_N\rangle}$, respectively.
All energies and rms radii are given, respectively, in units of MeV and fm.}
\label{tab:2}
\begin{center}
\begin{tabular}{ccccccccc}
\hline
\hline
\hspace{3mm}$N$\hspace{3mm} & \hspace{3mm}{nucleus}\hspace{3mm} 
     & \hspace{5mm}$\varepsilon$\hspace{5mm} & \hspace{5mm}$\langle t\rangle$\hspace{5mm}
     & \hspace{5mm}$\langle \upsilon_2\rangle$\hspace{5mm} & \hspace{5mm}$\langle \upsilon_C\rangle$\hspace{5mm}  
     & \hspace{5mm}$\langle \upsilon_{3}\rangle$\hspace{5mm}  & \hspace{5mm}$E$\hspace{5mm} 
     & \hspace{5mm}$\sqrt{\langle r^2_N\rangle}$\hspace{5mm} \\  
\hline
 3 & $^{12}$C  & 0.45 & 0.25 & $-1.90$ & 1.96 & 0.14 & 0.98 & 4.87 \\
 4 & $^{16}$O  & 0.76 & 0.27 & $-2.71$ & 2.86 & 0.33 & 1.84 & 5.23 \\
 5 & $^{20}$Ne & 1.12 & 0.27 & $-3.36$ & 3.68 & 0.53 & 3.04 & 5.55 \\
 6 & $^{24}$Mg & 1.51 & 0.26 & $-3.89$ & 4.44 & 0.70 & 4.63 & 5.85 \\
 7 & $^{28}$Si  & 1.91 & 0.26 & $-4.31$ & 5.13 & 0.84 & 6.61 & 6.13 \\
 8 & $^{32}$S   & 2.31 & 0.24 & $-4.64$ & 5.78 & 0.94 & 8.99 & 6.40 \\
 9 & $^{36}$Ar & 2.73 & 0.23 & $-4.89$ & 5.37 & 1.02 & 11.8 & 6.68 \\
10 & $^{40}$Ca & 3.13 & 0.22 & $-5.06$ & 6.91 & 1.60 & 15.0 & 6.95 \\
11 & $^{44}$Ti & 3.53 & 0.20 & $-5.15$ & 7.40 & 1.70 & 18.6 & 7.24 \\
\hline
\hline
\end{tabular}
\end{center}
\end{table}

\clearpage
\begin{table}
\caption{
Gaussian size parameter sets A and B used to obtain the deformed $N\alpha$ 
states with the axial symmetry by solving the Hill-Wheeler equation.
The unit of $b_1$ and $b_3$ is fm.
}
\label{tab:3}
\begin{center}
\begin{tabular}{ccccccc}
\hline
\hline
\hspace{5mm} set \hspace{5mm} 
                     & \hspace{5mm}$k_{max}$\hspace{5mm} & \hspace{5mm}$b_1^{(1)}$\hspace{5mm}
                     & \hspace{5mm}$b_1^{(k_{max})}$\hspace{5mm}
                     & \hspace{5mm}$K_{max}$\hspace{5mm} & \hspace{5mm}$b_3^{(1)}$\hspace{5mm} 
                     & \hspace{5mm}$b_3^{(K_{max})}$\hspace{5mm}  \\
\hline
A  & 12 & 0.4 &  9.0 & 12 & 0.5 & 9.5 \\
B  & 12 & 0.4 & 11.0 & 12 & 0.5 & 10.5 \\
\hline
\hline
\end{tabular}
\end{center}
\end{table}

\clearpage
\begin{table}
\caption{
Calculated total energy $E$, total kinetic energy $\langle T\rangle$, 
total $2\alpha$ nuclear potential energy $\langle V_2\rangle$, 
total Coulomb potential energy $\langle V_C\rangle$,
total $3\alpha$ potential energy $\langle V_3\rangle$ 
and nuclear rms radius $\sqrt{\langle r^2_N\rangle}$  
for each dilute $N\alpha$ state.
The effective energy of the $"2\alpha"$ system in the $N\alpha$ system
is denoted as $\varepsilon_2$ (see text).
All energies and rms radii are given in units of MeV and fm, respectively.
}
\label{tab:4}
\begin{center}
\begin{tabular}{ccccccccccc}
\hline
\hline
\multicolumn{9}{c}{deformed case} & \multicolumn{2}{c}{spherical case} \\
\hspace{2mm}{$N$}\hspace{2mm} & \hspace{2mm}nucleus\hspace{2mm}  & \hspace{3mm}{$\langle T\rangle$}\hspace{3mm} 
        & \hspace{3mm}{$\langle V_2\rangle$}\hspace{3mm} & \hspace{3mm}{$\langle V_C\rangle$}\hspace{3mm} 
        & \hspace{3mm}{$\langle V_3\rangle$}\hspace{3mm} 
        & \hspace{3mm}$E$\hspace{3mm} & ($\varepsilon_2$)
        & \hspace{3mm}$\sqrt{\langle r^2_N\rangle}$\hspace{3mm} 
        & \hspace{5mm}$E$\hspace{5mm} & \hspace{5mm}$\sqrt{\langle r^2_N\rangle}$\hspace{5mm} \\
\hline
 3 & $^{12}$C  & 2.24 & $-5.95$ & 3.62 & 0.09 & $-0.01$   & ($-0.00$)   & 3.73 & 0.64 & 4.18 \\
 4 & $^{16}$O  & 3.74 & $-11.31$ & 7.21 & 0.42 & $0.11$    & (0.02)       & 3.90 & 1.58 & 4.74 \\
 5 & $^{20}$Ne & 4.32 & $-15.92$ & 11.58 & 1.13 & $1.11$  & (0.11)       & 4.20 & 2.83 & 5.19 \\
 6 & $^{24}$Mg & 3.71 & $-18.78$ & 16.05 & 2.09 & $3.13$  & (0.21)        & 4.69 & 4.47 & 5.64 \\
 7 & $^{28}$S  & 3.13 & $-21.03$ & 20.58 & 0.09 & $5.58$   & (0.27)        & 5.24 & 6.48 & 5.99 \\
 8 & $^{32}$Si  & 2.77 & $-23.16$ & 25.31 & 0.09 & $8.30$   & (0.30)       & 5.79 & 8.93 & 6.33 \\
 9 & $^{36}$Ar  & 2.52 & $-25.11$ & 30.22 & 0.09 & $11.31$ & (0.31)       & 6.34 & 11.81 & 6.90 \\
10 & $^{40}$Ca & 2.35 & $-26.76$ & 35.21 & 0.09 & $14.62$ & (0.32)       & 6.90 & 14.98 & 7.26 \\
11 & $^{44}$Ti  & 2.26 & $-28.52$ & 40.58 & 0.09 & $18.27$  & (0.33)      & 7.38 & 18.53 & 7.54 \\
12 & $^{48}$Cr & 2.21 & $-30.43$ & 46.37 & 0.09 & $22.63$  & (0.34)      & 7.78 & 22.44 & 8.62 \\
\hline
\hline
\end{tabular}
\end{center}
\end{table}

\clearpage
\begin{table}
\caption{
Calculated energies $E$ of dilute $N\alpha$ states ($J^\pi=0^+$) together
with the nuclear rms radii $\sqrt{\langle r^2_N\rangle}$ and rms distances 
between $2\alpha$ bosons $\sqrt{\langle r^2_{\alpha\alpha}\rangle}$, where
we use the two Gaussian size parameter sets A and B, shown in Table~\ref{tab:3}.
The energy $E$ is measured from the respective $N\alpha$ threshold.
The units of energy and rms radius (distance) are MeV and fm,
respectively.
}
\label{tab:5}
\begin{center}
\begin{tabular}{cccccccc}
\hline
\hline
      &
      & \multicolumn{3}{c}{set A} 
      & \multicolumn{3}{c}{set B} \\
\hspace{3mm}$N$\hspace{3mm}  
      & \hspace{3mm}nucleus\hspace{3mm}
      & \hspace{5mm}$E$\hspace{5mm} 
      & \hspace{5mm}$\sqrt{\langle r^2_N\rangle}$\hspace{5mm}
      & \hspace{5mm}$\sqrt{\langle r^2_{\alpha\alpha}\rangle}$\hspace{5mm}
      & \hspace{5mm}$E$\hspace{5mm} 
      & \hspace{5mm}$\sqrt{\langle r^2_N\rangle}$\hspace{5mm}
      & \hspace{5mm}$\sqrt{\langle r^2_{\alpha\alpha}\rangle}$\hspace{5mm} \\
\hline
 3  &  $^{12}$C    & $-0.01$ & 3.73 & 5.73 & $-0.01$ & 3.72 & 5.72 \\
 4  &  $^{16}$O    &   0.11   & 3.90  & 5.72 & 0.11     & 3.90  & 5.72 \\
 5  &  $^{20}$Ne   &  1.11    & 4.20  & 6.06 & 1.13    & 4.20  & 6.06 \\
 6  &  $^{24}$Mg   &  3.13    & 4.69  & 6.76 & 3.13    & 4.68   & 6.75 \\
 7  &  $^{28}$Si    &  5.58     & 5.24  & 7.57 & 5.59    & 5.24  & 7.57 \\
 8  &  $^{32}$S     &  8.30    & 5.79  & 8.36 & 8.30    & 5.78   & 8.35  \\
 9  &  $^{36}$Ar    & 11.31   & 6.34  & 9.16 & 11.31   & 6.35  & 9.17 \\
 10 &  $^{40}$Ca   & 14.62    & 6.90  & 9.97 & 14.61   & 7.03  & 10.58 \\
 11 &  $^{44}$Ti    & 18.27    & 7.38  & 10.65 &18.22   & 7.89  & 11.42 \\
 12 &  $^{48}$Cr   & 22.27    & 7.78  & 11.21 & 22.10  & 8.81 & 12.77 \\
\hline
\hline
\end{tabular}
\end{center}
\end{table}

\clearpage

\clearpage

\clearpage
\begin{figure}
\begin{center}
\begin{minipage}{0.90\linewidth}
\includegraphics*[width=\linewidth]{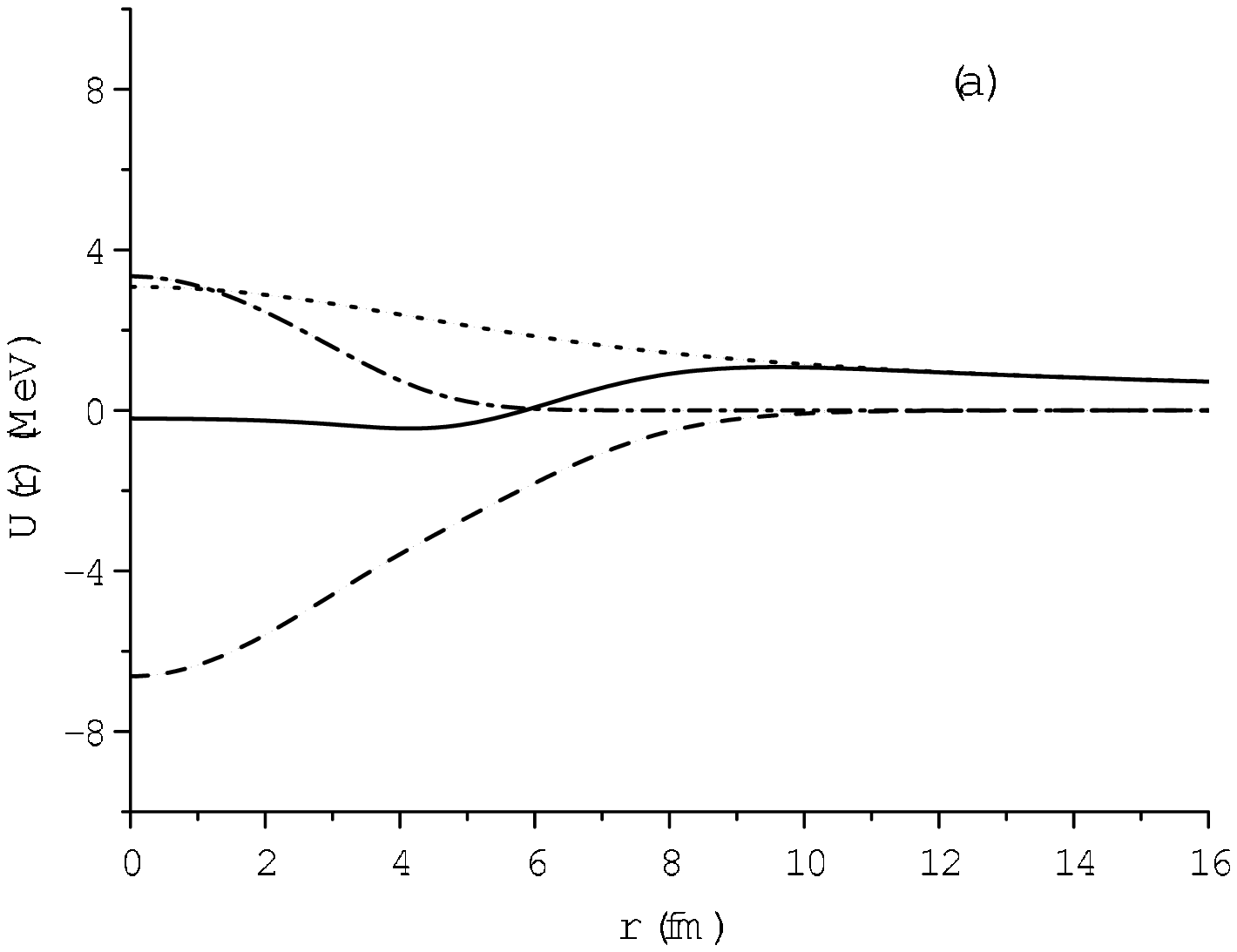}
\end{minipage}
\begin{minipage}{0.90\linewidth}
\includegraphics*[width=\linewidth]{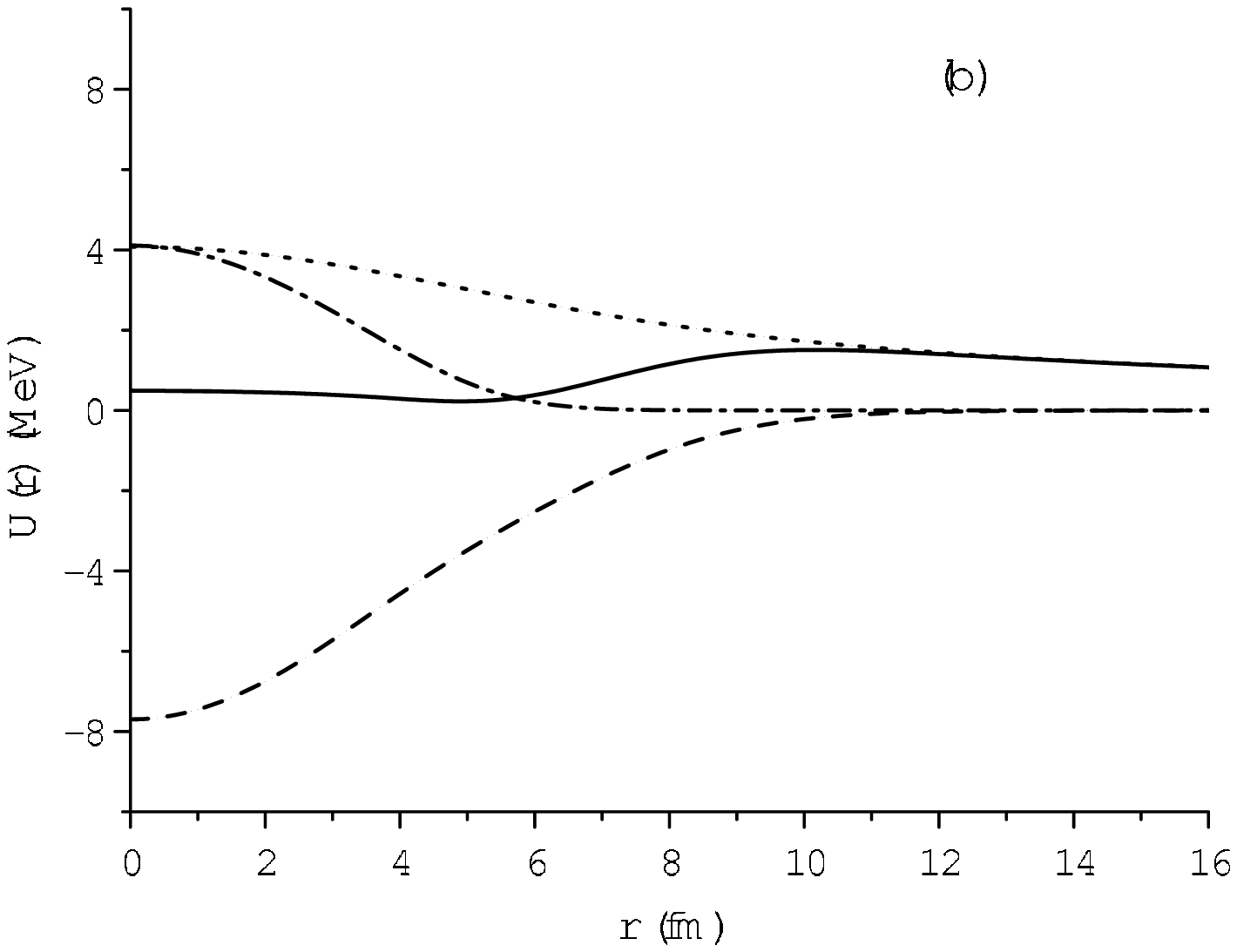}
\end{minipage}
\end{center}
\end{figure}

\begin{figure}
\begin{center}
\begin{minipage}{0.90\linewidth}
\includegraphics*[width=\linewidth]{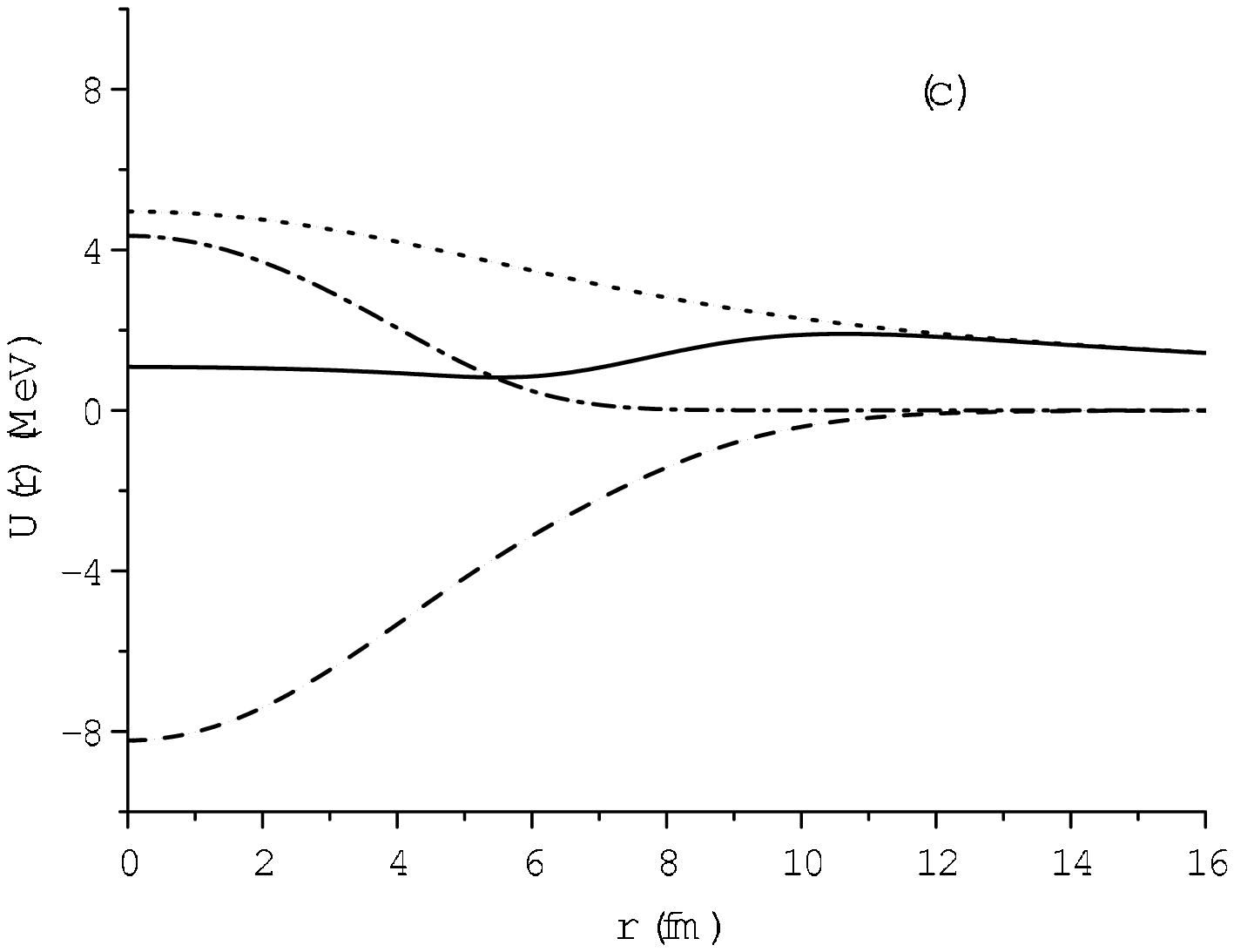}
\end{minipage}
\begin{minipage}{0.90\linewidth}
\includegraphics*[width=\linewidth]{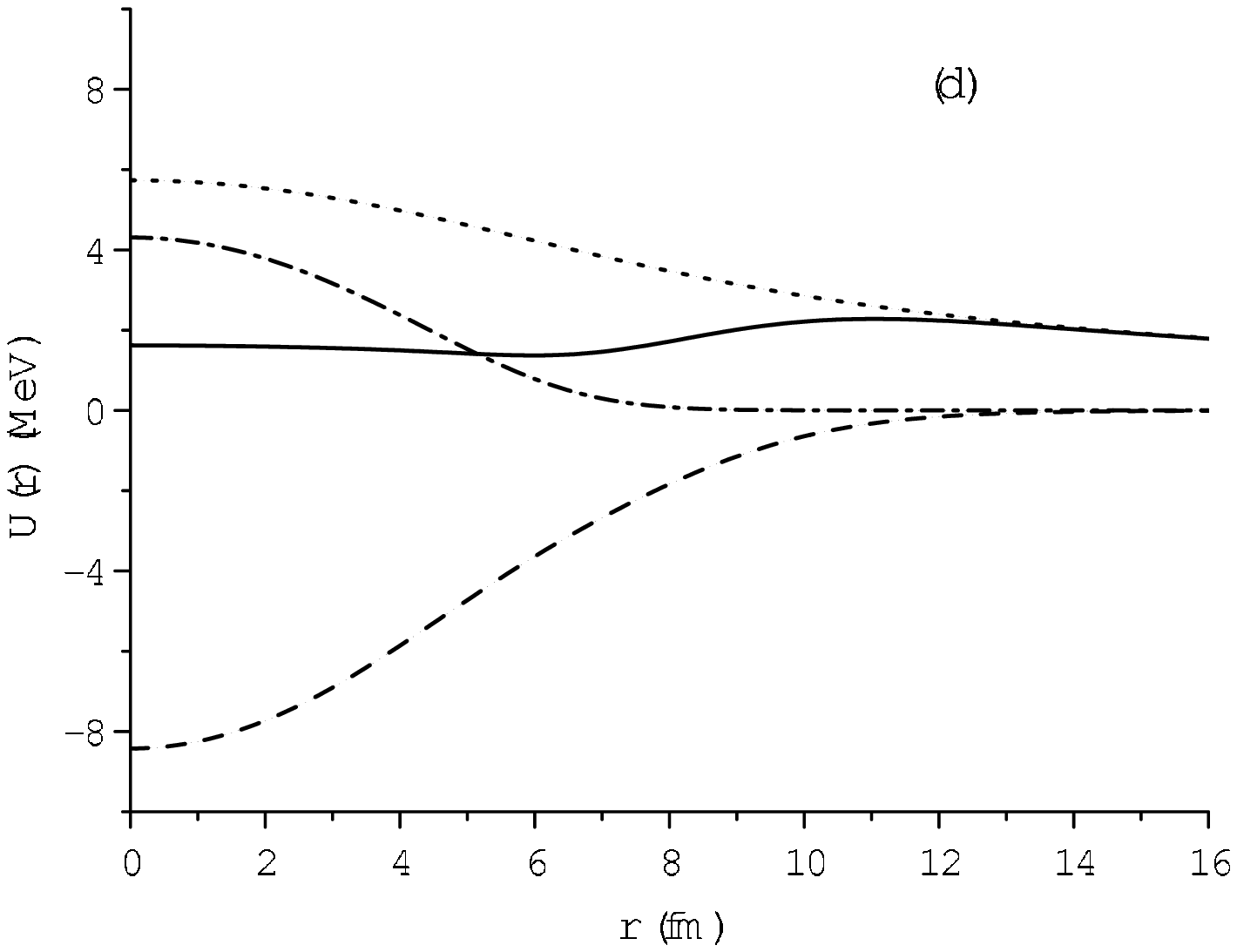}
\end{minipage}
\end{center}
\end{figure}

\begin{figure}
\begin{center}
\begin{minipage}{0.90\linewidth}
\includegraphics*[width=\linewidth]{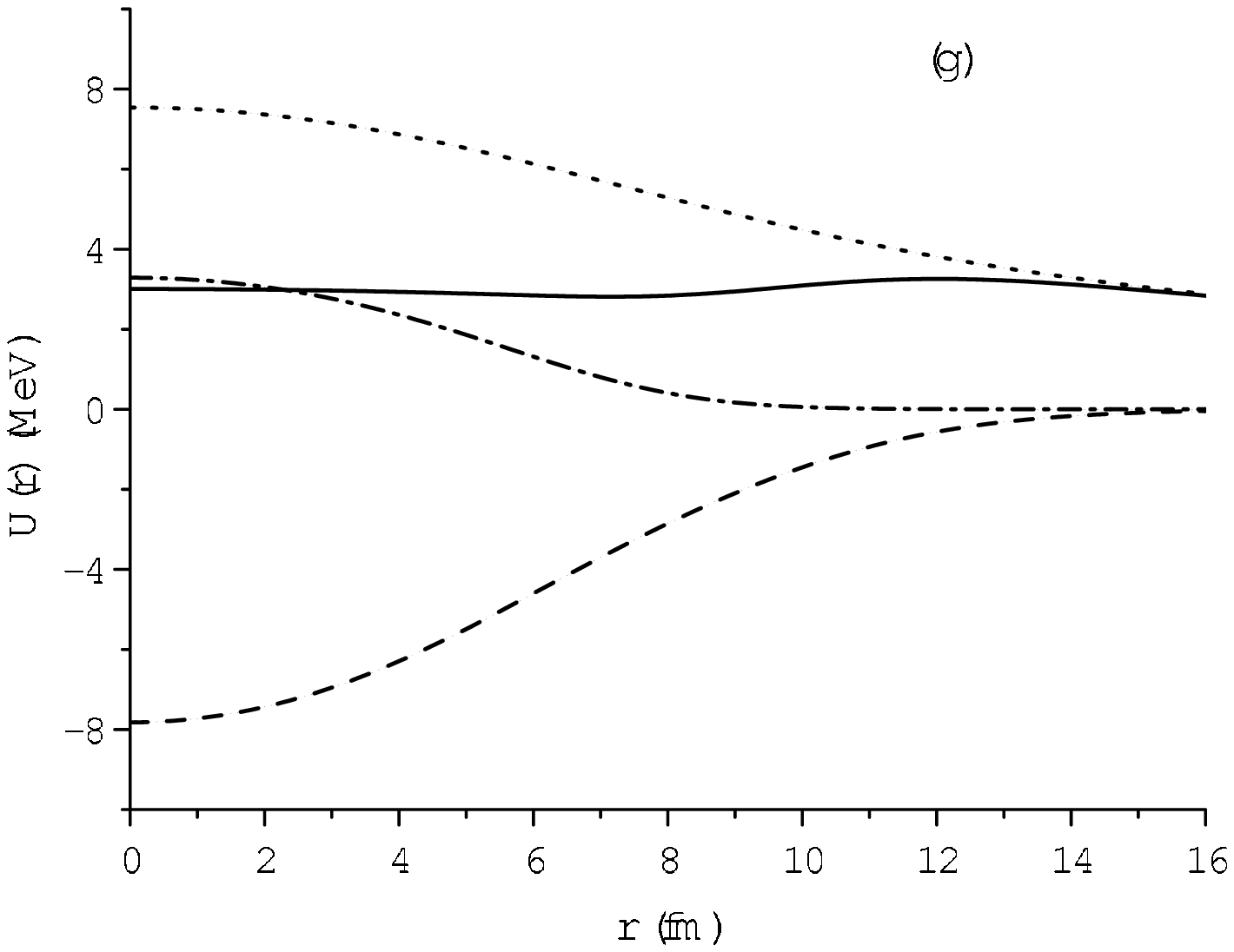}
\end{minipage}
\begin{minipage}{0.90\linewidth}
\includegraphics*[width=\linewidth]{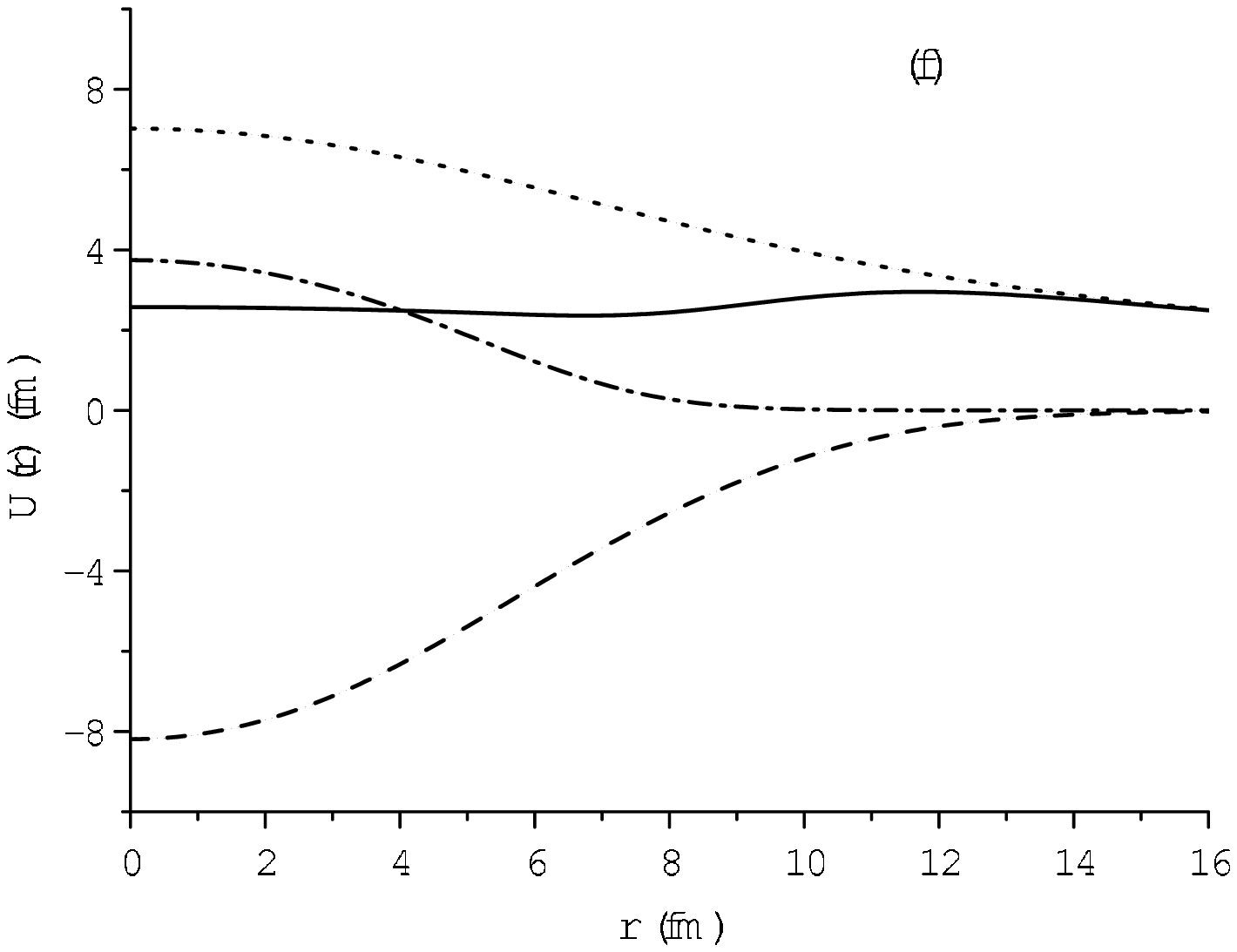}
\end{minipage}
\end{center}
\end{figure}

\begin{figure}
\begin{center}
\begin{minipage}{0.90\linewidth}
\includegraphics*[width=\linewidth]{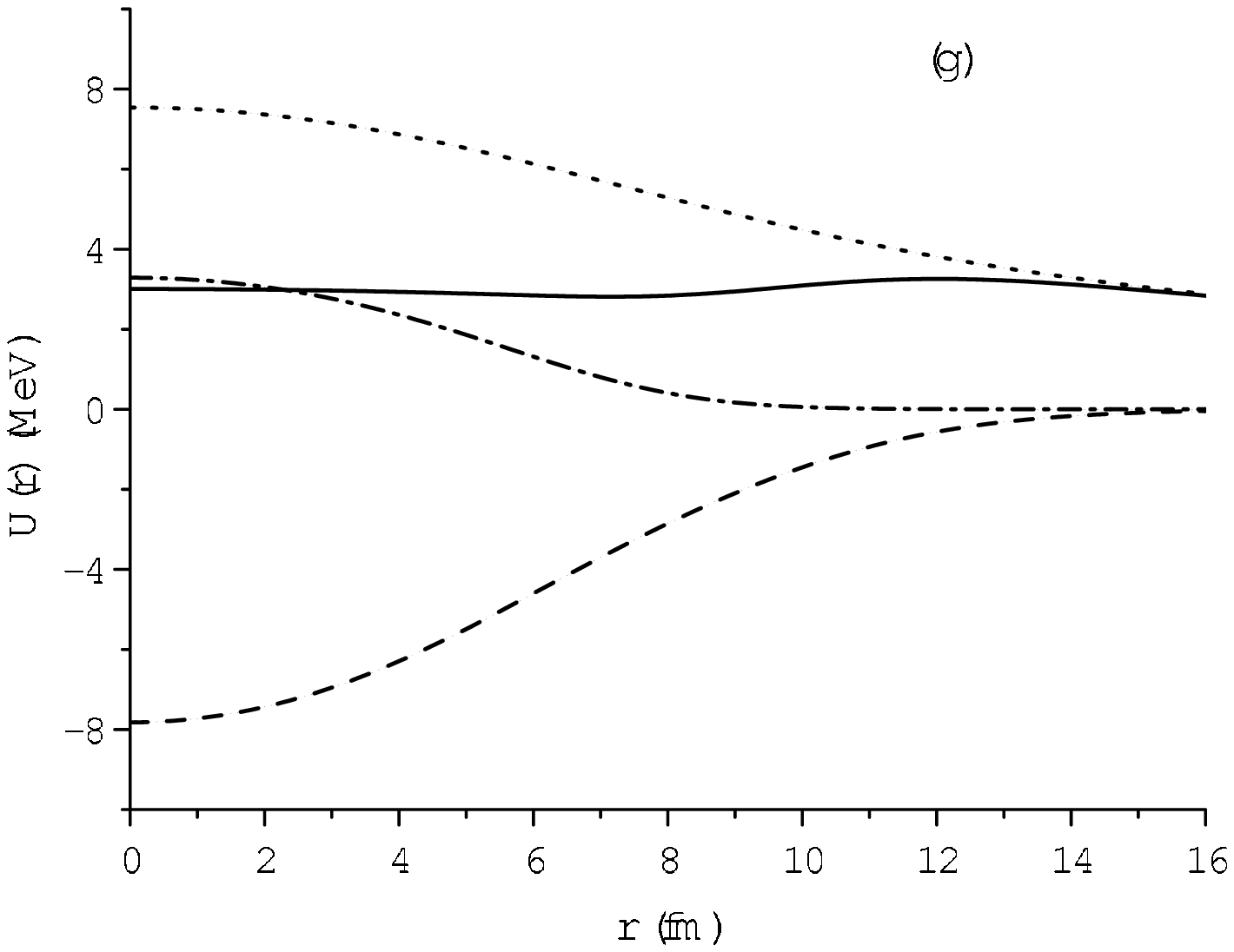}
\end{minipage}
\begin{minipage}{0.90\linewidth}
\includegraphics*[width=\linewidth]{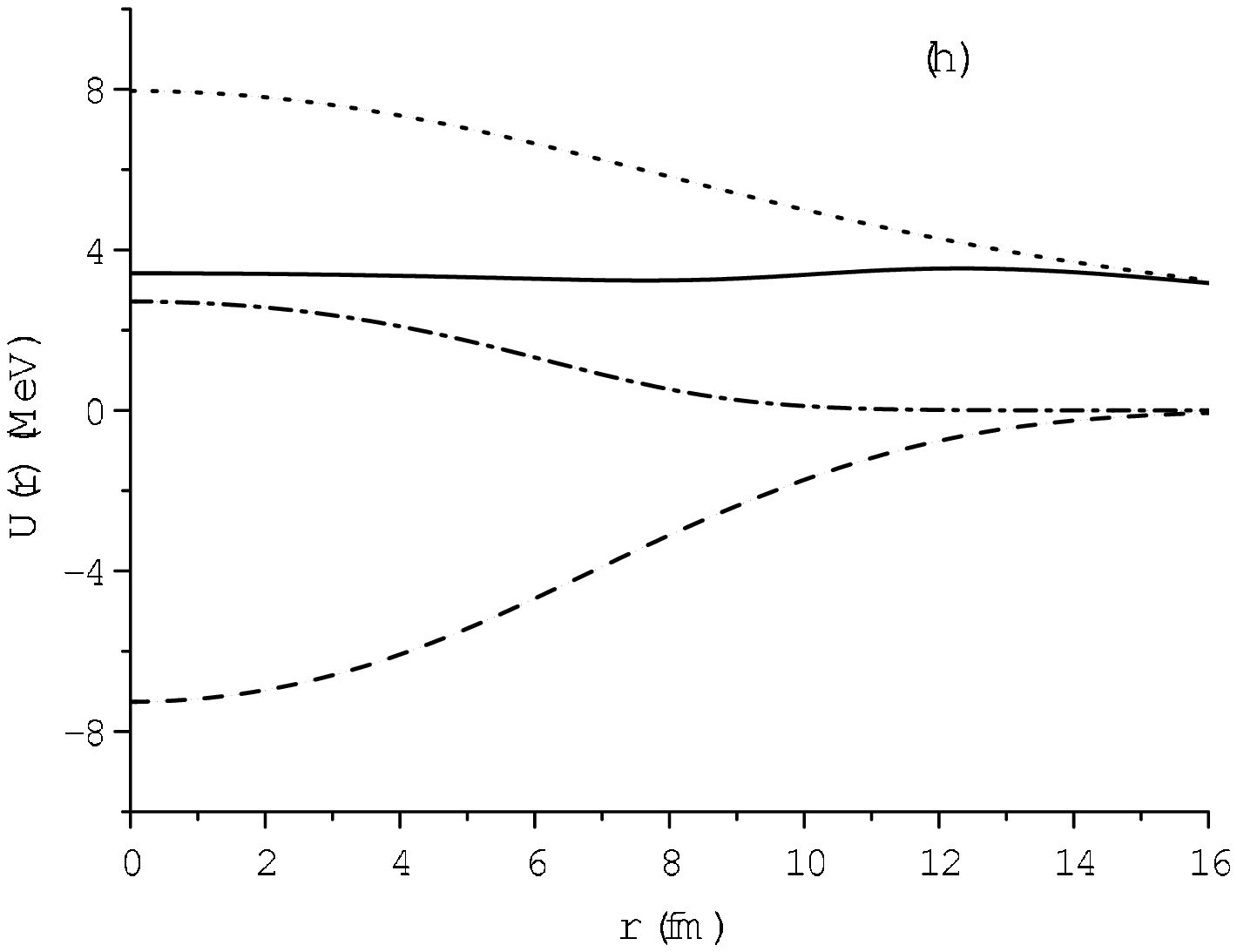}
\end{minipage}
\end{center}
\end{figure}

\clearpage

\clearpage
\begin{figure}
\begin{center}
\begin{minipage}{0.90\linewidth}
\includegraphics*[width=\linewidth]{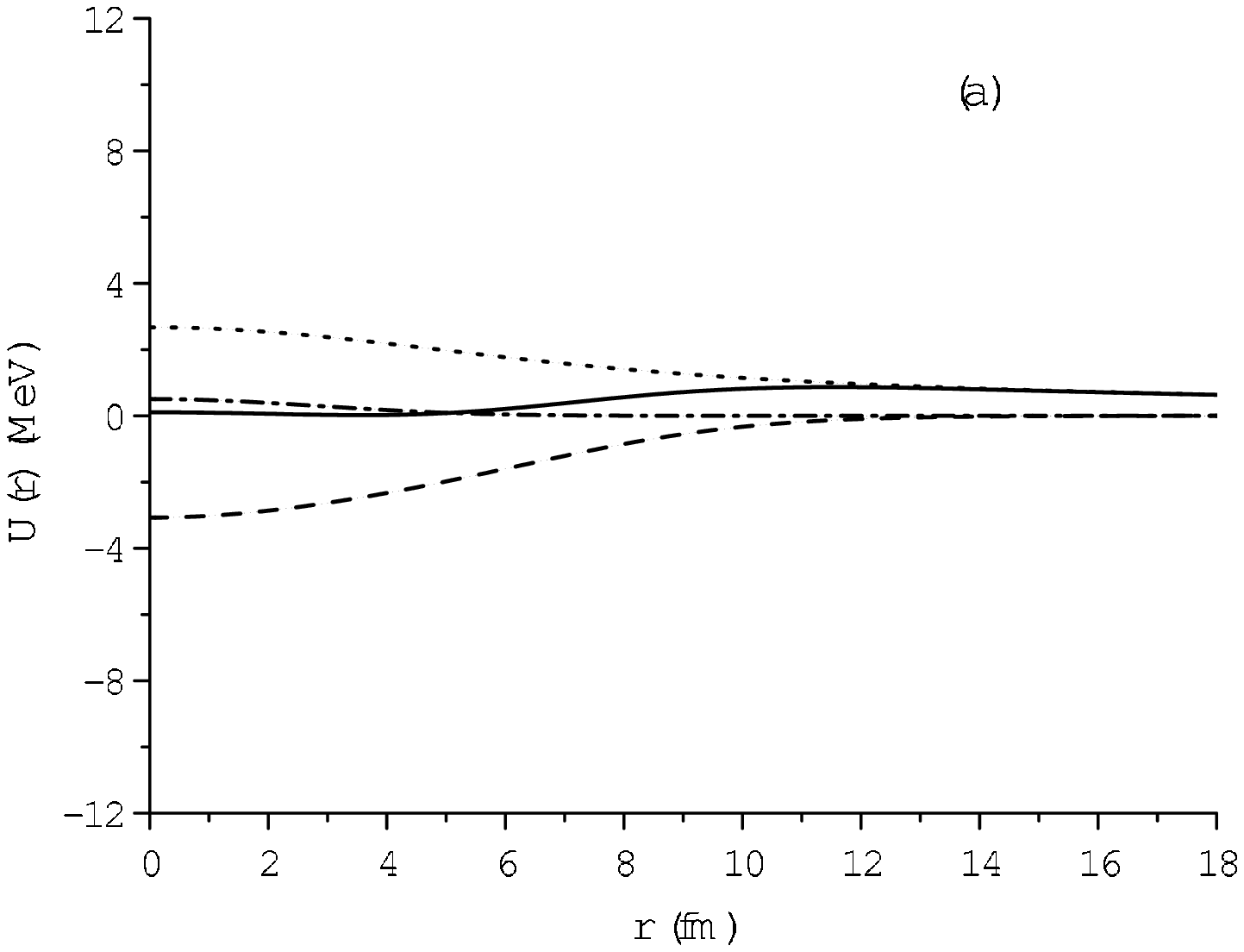}
\end{minipage}
\begin{minipage}{0.90\linewidth}
\includegraphics*[width=\linewidth]{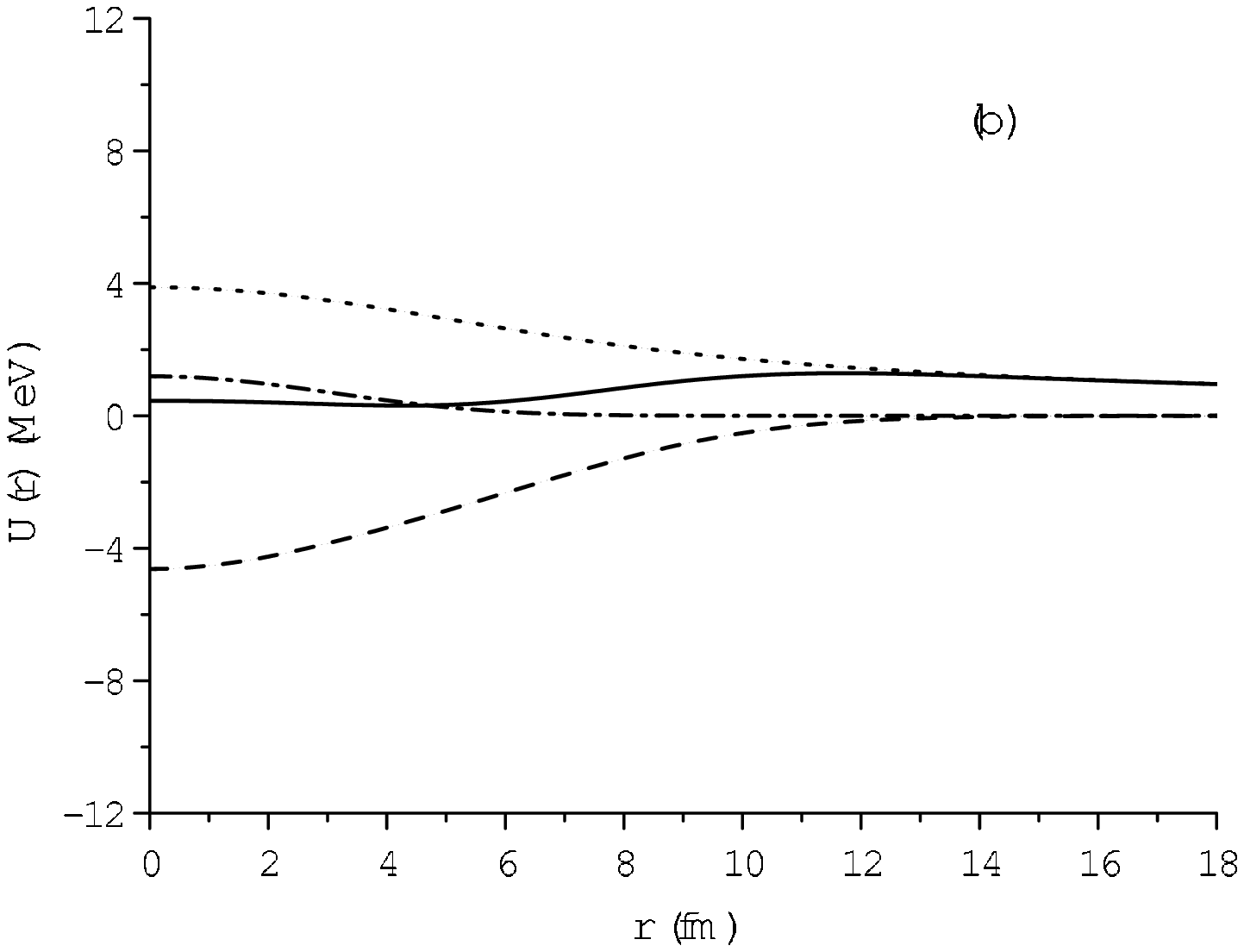}
\end{minipage}
\end{center}
\end{figure}

\begin{figure}
\begin{center}
\begin{minipage}{0.90\linewidth}
\includegraphics*[width=\linewidth]{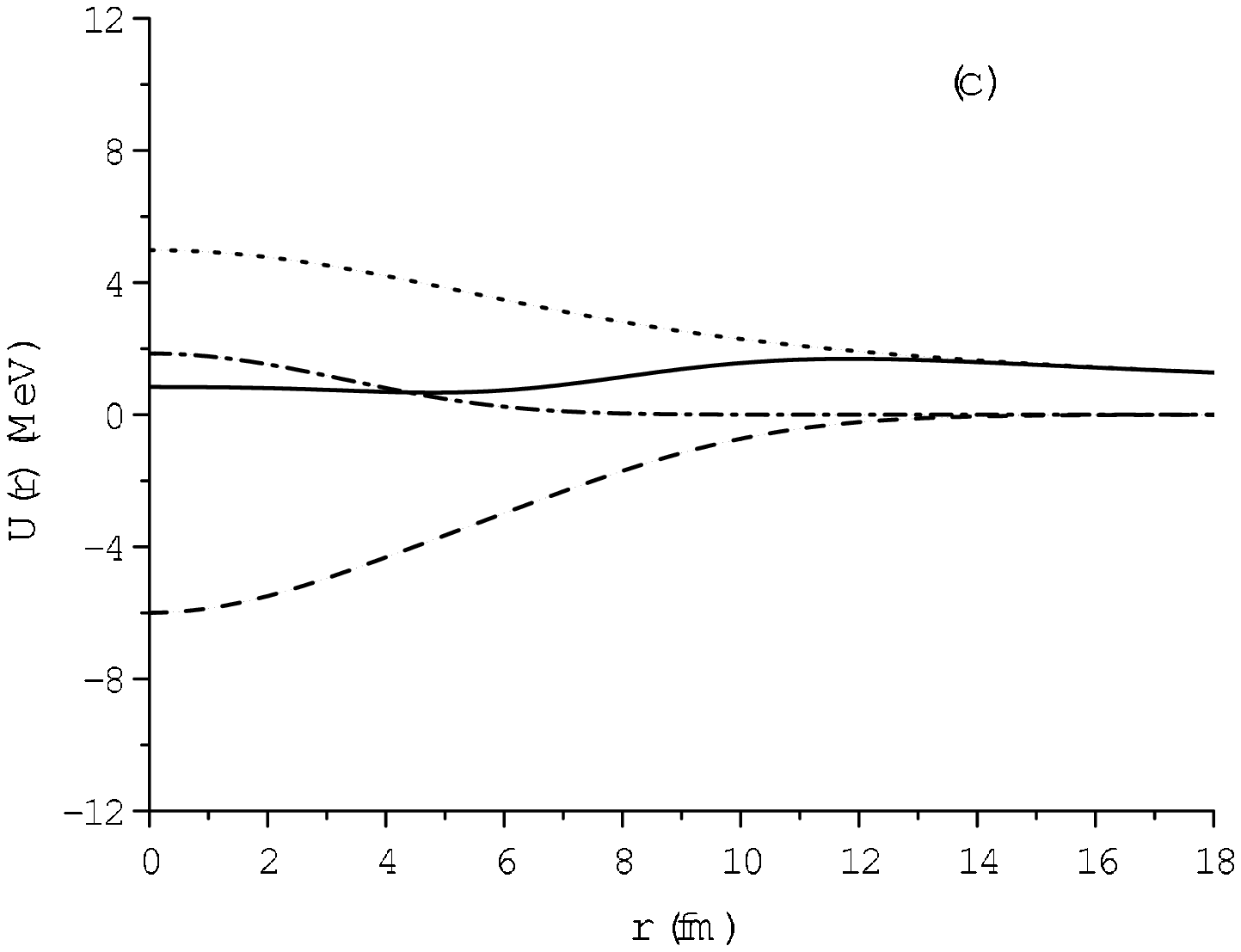}
\end{minipage}
\begin{minipage}{0.90\linewidth}
\includegraphics*[width=\linewidth]{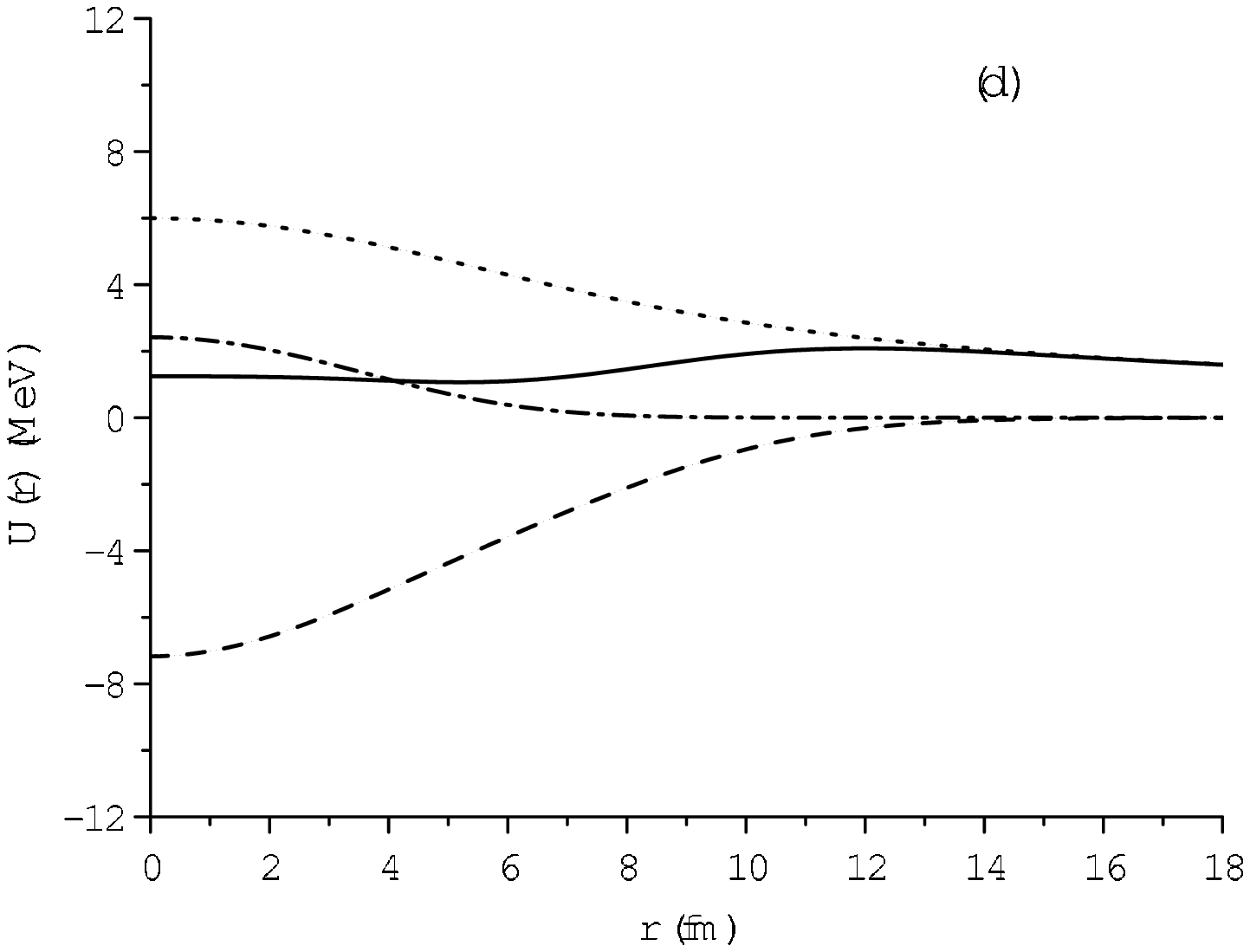}
\end{minipage}
\end{center}
\end{figure}

\begin{figure}
\begin{center}
\begin{minipage}{0.90\linewidth}
\includegraphics*[width=\linewidth]{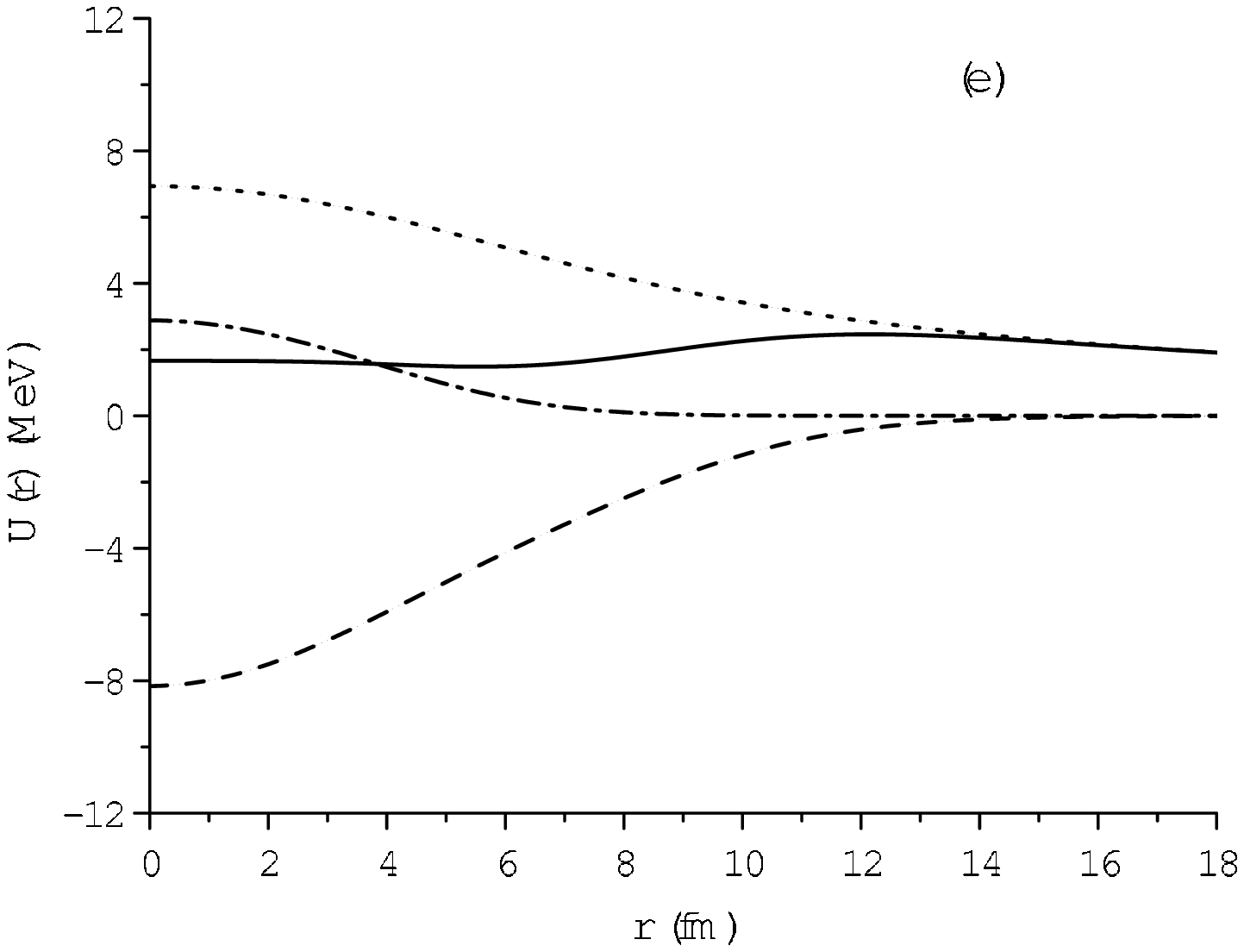}
\end{minipage}
\begin{minipage}{0.90\linewidth}
\includegraphics*[width=\linewidth]{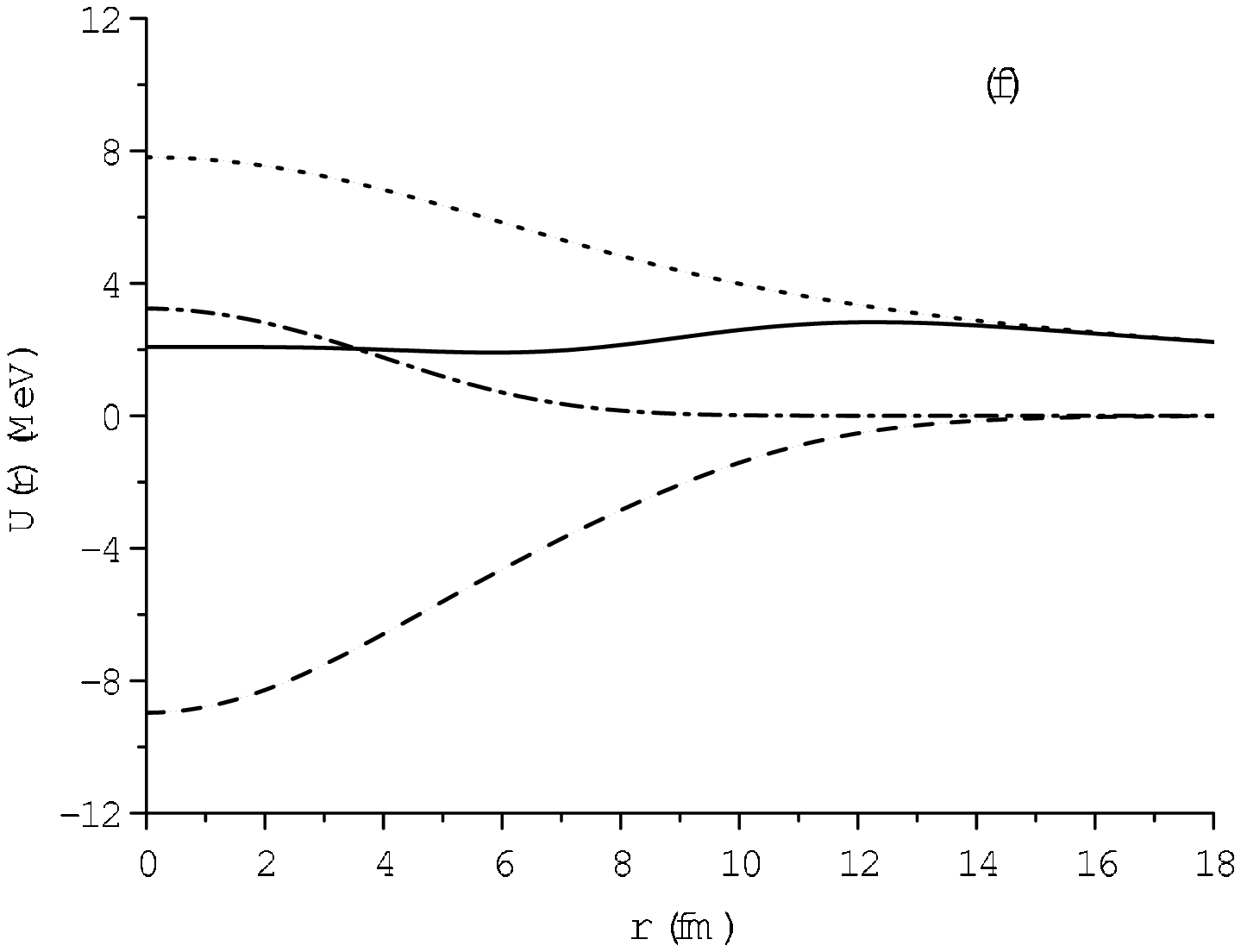}
\end{minipage}
\end{center}
\end{figure}

\begin{figure}
\begin{center}
\begin{minipage}{0.90\linewidth}
\includegraphics*[width=\linewidth]{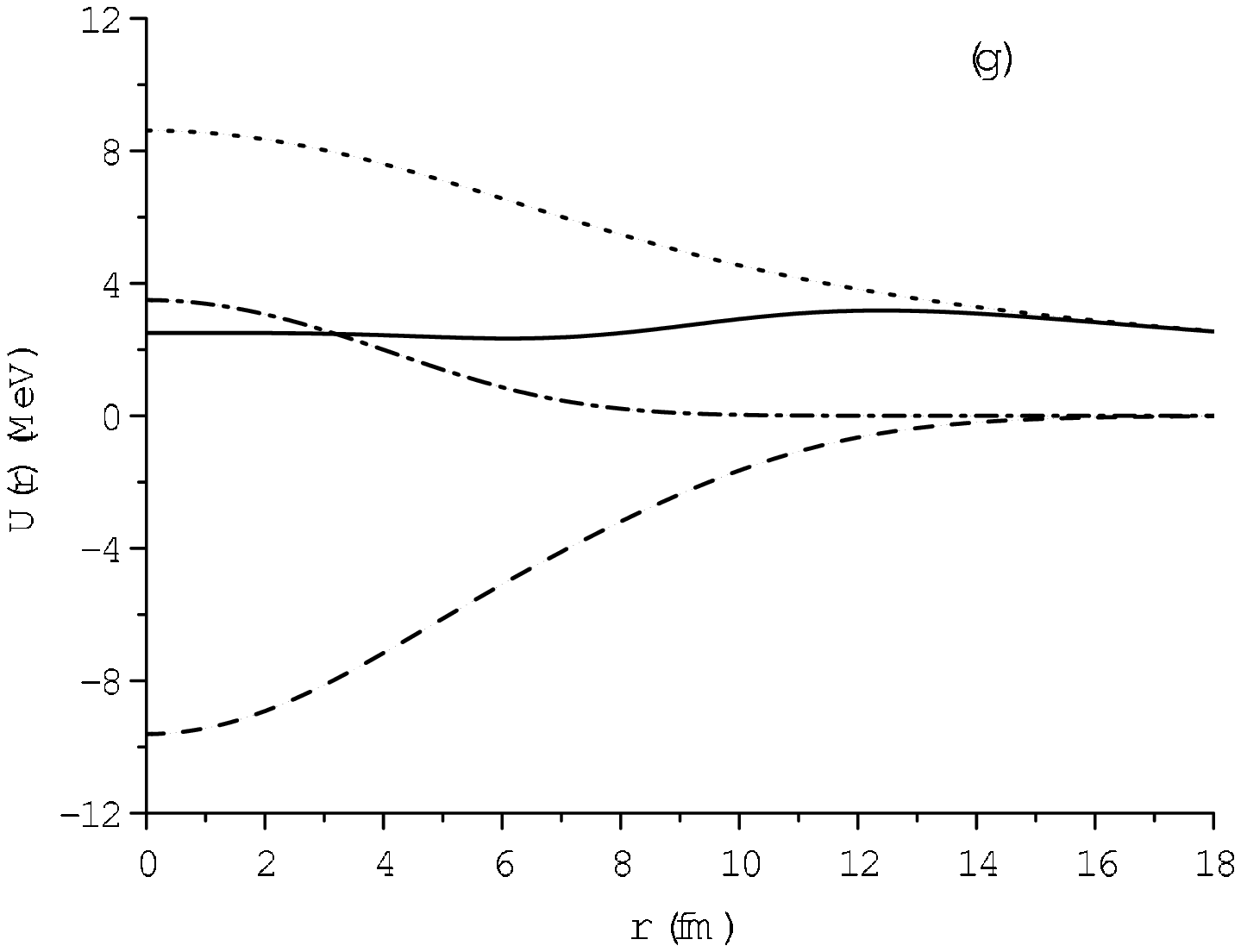}
\end{minipage}
\begin{minipage}{0.90\linewidth}
\includegraphics*[width=\linewidth]{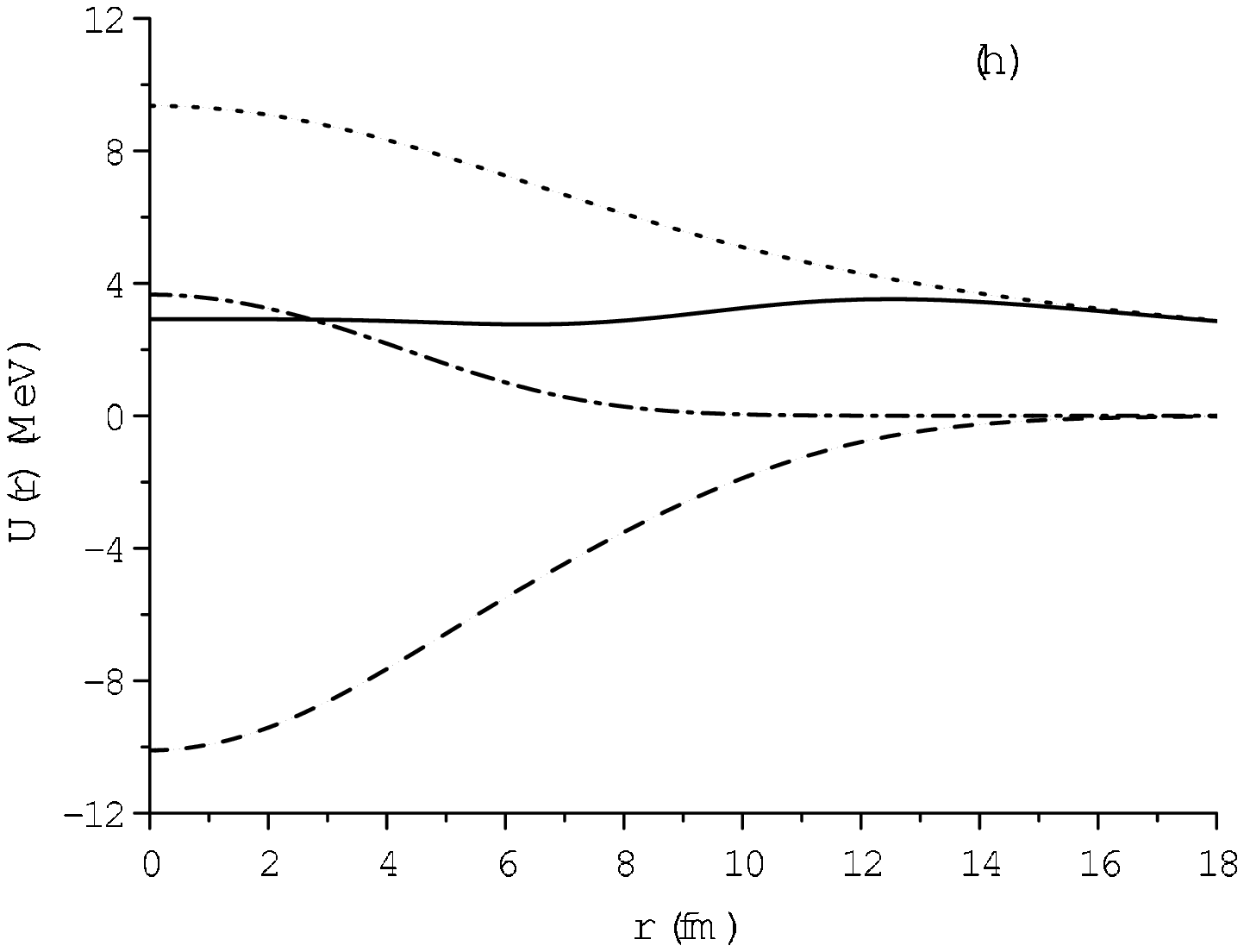}
\end{minipage}
\end{center}
\end{figure}

\clearpage

\clearpage

\clearpage


\clearpage

\begin{thebibliography}{99}

\bibitem{Wildermuth77}
   K.~Wildermuth and Y.C.~Tang, {\it A Unified Theory of the Nucleus} 
   (Vieweg, Braunschweig, Germany, 1977).

\bibitem{Brink66}
   D.M.~Brink, in {\it Many-Body Description of Nuclear Structure and Reactions}, 
   Proceedings of the International School of Physics "Enrico Fermi", Course 36,
   edited by C.~Bloch (Academic Press, New York, 1966).

\bibitem{Bertsch71}
   G.F. Bertsch and W. Bertozzi, Nucl.\ Phys.\ {\bf A165}, 199 (1971).

\bibitem{Fujiwara80}
   Y.~Fujiwara, H.~Horiuchi, K.~Ikeda, M.~Kamimura, K.~Kato, Y.~Suzuki, and E.~Uegaki,
   Prog.\ Theor.\ Phys.\ Suppl.\ No.\ 68, 29 (1980). 

\bibitem{Ikeda68}
   K.~Ikeda, N.~Takigawa, and H.~Horiuchi, Prog.\ Theor.\ Phys.\ Suppl.\ Extra\ Number, 464 (1968). 

\bibitem{Horiuchi86}
   For example, H.~Horiuchi and K.~Ikeda, {\it Cluster Model of the Nucleus},
   International Review of Nuclear Physics, World Scientific Publishing Co., {\bf 4}, 1 (1986).

\bibitem{Dalfovo99}
   F.~Dalfovo, S.~Giorgini, L.P.~Pitaevskii, and S.~Stringari, Rev.\ Mod.\ Phys.\ 
   {\bf 71}, 463 (1999).

\bibitem{Ropke98}
   G.~R\"opke, A.~Schnell, P.~Schuck, and P.~Nozieres, 
   Phys.\ Rev.\ Lett.\ {\bf 80}, 3177 (1998).     

\bibitem{Beyer00}
   M.~Beyer, S.A.~Sofianos, C.~Kuhrts,  G.~R\"opke, and P.~Schuck, 
   Phys.\ Lett.\ B {\bf 80}, 247 (2000).     

\bibitem{Tohsaki01} 
   A.~Tohsaki, H.~Horiuchi, P.~Schuck and G.~R\"opke, 
   Phys.\ Rev.\ Lett.\ {\bf 87}, 192501 (2001). 

\bibitem{Funaki02}
   Y.~Funaki, H.~Horiuchi, A.~Tohsaki, P.~Schuck and G.~R\"opke, 
   Prog.\ Theor.\ Phys.\ {\bf 108}, 297 (2002).

\bibitem{Pitaevskii61}
   L.P.~Pitaevskii, Zh.\ Eksp.\ Theor.\ Fiz.\ {\bf 40}, 646 (1961) [Sov.\ Phys.\ JETP {\bf 13}, 451 (1961)];
   E.P.~Gross, Nuovo\ Cimento {\bf 20}, 454 (1961); J.\ Math.\ Phys.\ {\bf 4}, 195 (1963).

\bibitem{Hill53}
   D.L.~Hill and J.A. Wheeler, Phys.\ Rev.\ {\bf 89}, 1102 (1953).\\
   J.J.~Griffin and J.A.~Wheeler, Phys.\ Rev.\ {\bf 108}, 311 (1957).

\bibitem{Ali66}
   S.~Ali and A.R.~Bodmer, Nucl.\ Phys.\ {\bf 80}, 99 (1966).

\bibitem{Tamagaki77}
   R.~Tamagaki and Y.~Fujiwara,  Prog.\ Theor.\ Phys.\ Suppl.\ No.\ 61, 229 (1977).

\bibitem{Suzuki02}
   Y.~Suzuki and M.~Takahashi, Phys.\ Rev.\ C\ {\bf 65}, 064318 (2002).

\bibitem{footnote}
   In Ref.~\cite{Suzuki02}, the calculated value is 2.64 fm, which is not taken 
   into account the finite size effect of the $\alpha$ particle. With help of 
   the Eq.~(\ref{HW_rms_N}) in this paper, the nuclear rms radius 
   $\sqrt{\langle r^2_N\rangle}$ is 3.15 fm. 

%
\bibitem{Fukatsu92}
   K.~Fukatsu and K.~Kato, Prog.\ Theor.\ Phys.\ {\bf 87}, 151 (1992). 

\bibitem{Saito68}
   S.~Saito, Prog.\ Theor.\ Phys.\ {\bf 40} (1968) 893; {\bf 41}, 705 (1969). 

\bibitem{Ajzenberg90}
   F.~Ajzenberg-Selove, Nucl.\ Phys.\ {A506}, 1 (1990).

\end{thebibliography}
\end{document}